\documentclass[journal]{IEEEtran}
\usepackage{cite}

\ifCLASSINFOpdf
   \usepackage[pdftex]{graphicx}
\else
\fi
\usepackage{amsmath}
\usepackage{algorithm}
\usepackage{algorithmic}

\usepackage{array}

\ifCLASSOPTIONcompsoc
  \usepackage[caption=false,font=normalsize,labelfont=sf,textfont=sf]{subfig}
\else
  \usepackage[caption=false,font=footnotesize]{subfig}
\usepackage{url}

\usepackage{amsfonts}
\usepackage{comment,soul,bm}
\usepackage{enumerate,enumitem,tabularx,multirow}
\usepackage[dvipsnames]{xcolor}
\newcommand{\rev}[1]{\textcolor{Green}{#1}}
\newcommand{\revb}[1]{\textcolor{blue}{#1}}

\DeclareMathOperator*{\argmin}{arg\, min}

\begin{document}
%
\title{Dynamic distributed decision-making for resilient resource reallocation in disrupted manufacturing systems
\thanks{This work was funded in part by NSF 1544678}}
%
%
%

\author{Mingjie Bi, Ilya Kovalenko, Dawn M. Tilbury, Kira Barton
\thanks{Mingjie Bi is with the Robotics Institute, 
        University of Michigan, Ann Arbor, MI 48109, USA
        {\tt\small mingjieb@umich.edu}}
\thanks{Ilya Kovalenko is with the Department of Mechanical Engineering and Industrial \& Manufacturing Engineering, 
         Pennsylvania State University, University Park, PA 16802, USA
        {\tt\small iqk5135@psu.edu}}
\thanks{Dawn M. Tilbury and Kira Barton are with the Department of Mechanical Engineering and the Robotics Institute, 
        University of Michigan, Ann Arbor, MI 48109, USA
        {\tt\small \{tilbury, bartonkl\}@umich.edu}}
}
%
%

\markboth{IEEE Journal Template}{}
%



\maketitle


\begin{abstract}
The COVID-19 pandemic brings many unexpected disruptions, such as frequently shifting markets and limited human workforce, to manufacturers. 
To stay competitive, flexible and real-time manufacturing decision-making strategies are needed to deal with such highly dynamic manufacturing environments.
One essential problem is dynamic resource allocation to complete production tasks, especially when a resource disruption (e.g. machine breakdown) occurs.
Though multi-agent methods have been proposed to solve the problem in a flexible and agile manner, the agent internal decision-making process and resource uncertainties have rarely been studied.
This work introduces a model-based resource agent (RA) architecture that enables effective agent coordination and dynamic agent decision-making.
Based on the RA architecture, a rescheduling strategy that incorporates risk assessment via a clustering agent coordination strategy is also proposed.
A simulation-based case study is implemented to demonstrate dynamic rescheduling using the proposed multi-agent framework.
The results show that the proposed method reduces the computational efforts while losing some throughput optimality compared to the centralized method.
Furthermore, the case study illustrates that incorporating risk assessment into rescheduling decision-making improves the throughput.
\end{abstract}


%
\IEEEpeerreviewmaketitle


\section{Introduction}
\label{sec:introduction}

Due to COVID-19, manufacturing enterprises have faced unprecedented challenges, including emerging large-volume demands for everyday items, new medical device production requirements, and limited human workforce.
Thus, the need to deal with dynamic manufacturing environments that include frequently shifting product demands, the customization of products, or unexpected disruptions (e.g. machine breakdowns) has been highlighted by the pandemic~\cite{kumar2020covid, bi2022model}.
To stay competitive, manufacturing enterprises must develop flexible and real-time decision-making strategies to adapt to this dynamic environment~\cite{li2020intelligent}.
One research area that focuses on dynamic decision-making is the effective allocation of limited resources through a flexible response to unexpected disruptions using control and coordination of components on the shop floor.

Currently, most manufacturing systems apply centralized decision-making strategies, such as mathematical programming~\cite{abumaizar1997rescheduling} and reinforcement learning~\cite{lee2022reinforcement,liu2022deep,park2019reinforcement,yang2021intelligent,palombarini2021end}, to generate optimal resource allocation plans based on specific objectives (e.g., cost and throughput).
However, since a centralized decision-making process includes the consideration of the entire system through a global view, it generally requires significant computational efforts to calculate.
Therefore, systems with centralized strategies often lack the ability to respond dynamically and quickly to unexpected disruptions~\cite{leitao2009agent}.
To improve the flexibility and agility of these systems, distributed decision-making strategies, where multiple system entities interact collaboratively to make decisions, have been proposed in several studies~\cite{leitao2009agent,shen2006agent,kovalenko2022toward}.
\begin{figure}[!t]
\centering
\includegraphics[width=1\columnwidth]{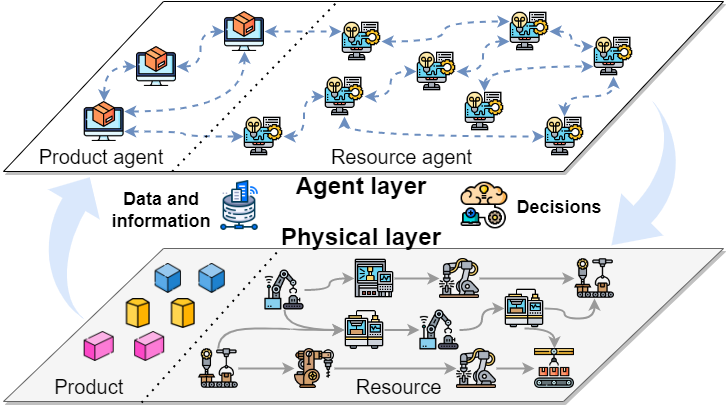}
\caption{An example of multi-agent manufacturing system}
\label{fig:mams}
\end{figure}

One type of distributed strategy that has been proposed to enable dynamic decision-making is multi-agent control~\cite{leitao2009agent,bi2021dynamic}.
A multi-agent system consists of various autonomous agents performing coordination and decision-making~\cite{wooldridge2009introduction}. 
In the manufacturing domain, product agents (PAs) and resource agents (RAs) have been described in most existing multi-agent architectures~\cite{kovalenko2019dynamic}.
A PA is responsible for fulfilling production requirements for its associated physical part through interactions with other agents, while an RA provides high-level control for its associated resource in the physical layer~\cite{kovalenko2019model}.
Through the coordination and decision-making of PAs and RAs, flexibility and responsiveness in manufacturing systems can be improved~\cite{leitao2009agent}.

The initial consideration of a dynamic response within the resource reallocation problem has been studied in the authors' previous work~\cite{bi2021dynamic}.
This previous work proposed a dynamic resource reallocation strategy based on RA capability clustering to improve the reallocation performance.
However, this approach did not consider resiliency or robustness, which refers to the ability to recover system performance under different environmental conditions such as uncertainties~\cite{mejia2020robust}.
In many systems, the quality and execution of a rescheduling solution lie in the resiliency and robustness of the new schedule.
Thus, to ensure a dynamic and resilient response, the decision-making process must incorporate uncertainty and potential risk factors into the optimization.
Furthermore, to enable effective communication and enhanced decision-making, an RA should understand its own objectives and the complex manufacturing environment.
However, most existing multi-agent frameworks do not incorporate this information into the RA architecture nor do they consider uncertainties and risks within the rescheduling problem~\cite{zhang2017flexible, rehberger2016agent,wong2006integrated,rodrigues2018decentralized}.



To address these limitations, this work builds on the agent model and rescheduling strategy proposed in~\cite{bi2021dynamic}. 
The contributions of this paper over the previous work include: (1) the extension and generalization of an RA architecture that includes a Knowledge Base, a Communication Manager, and a Decision Manager, (2) the development of a risk assessment approach and a dynamic and resilient resource reallocation strategy, and (3) an evaluation of manufacturing system performance when implementing the proposed approach within a simulated manufacturing facility.

The rest of the paper is organized as follows.
Background regarding the decision-making for rescheduling is discussed in Section~\ref{sec:background}.
Section~\ref{sec:problemformulation} describes the agents and resource allocation problem formulation.
The RA architecture, including the Knowledge Base, Communication Manager, and Decision Manager, is presented in Section~\ref{sec:RAarchitecture}.
Section~\ref{sec:reschedulingprocess} describes the resource reallocation process via RA coordination. 
In Section~\ref{sec:casestudy}, a simulation case study with the proposed architecture is provided, and conclusions are in Section~\ref{sec:conclusion}.

\section{Background}
\label{sec:background}
\begin{table*}
\centering
\caption{Nomenclature for the RA architecture}
\begin{tabular}{lcl} 
 \hline
 Production schedule &  &\\
 \hline
 $S$ &  & Function that maps agents to the product and resource schedule\\
 $s$ &  & Function that maps each agent to a sequence of events in the schedule of the agent\\
 $Ag$ &  & Function that maps events to particular agents\\
 $T_s$ &  & Function that maps event sequence to the start and end times of each event\\
 \hline
 Agents &  &\\
 \hline
 $X$ & &  Set of states of a product\\
 $E$ & &  Set of events representing operations that change product state\\
 $T_r$ &  & State transition function representing how events drive state changes\\
 $x_i=(x^\ell_i,x^c_i)$  & & State of a product that describes its location and physical composition\\
 $T$  & & Cost function for performing events\\
 $A_t$  & & Function that maps events to resource attributes\\
 $P_q$  & & Set of production requirements for scheduled events\\
 $C_{\ell}$  & & Function that maps events to clustering RAs\\
 \hline
 Rescheduling process  & &\\
 \hline
 $RA_d$  & & The disrupted RA\\
 $E_d$& & Sequence of affected events in the resource schedule of $RA_d$\\
 $s_d$  & & Sequence of events that need to be replaced\\
 $x_{prior}$  & & State before the event sequence that should be replaced ($s_d$) in the initial product schedule\\
 $x_{post}$  & & State after the event sequence that should be replaced ($s_d$) in the initial product schedule\\
 $s_{new}$  & & Sequence of events that can replace the event sequence $s_d$\\
 
 $H$  & & Function that calculates earliest available time for a resource to perform an event \\
 $R$  & & Function that calculate risk of a new event sequence\\
 \hline
\end{tabular}
\label{tab:nomenclature}
\end{table*}





The scheduling and rescheduling problems have been widely studied via centralized decision-making, such as mathematical programming~\cite{fu2020heterogeneous,fu2022robust} and reinforcement learning~\cite{lee2022reinforcement,liu2022deep,park2019reinforcement,yang2021intelligent,palombarini2021end}.
However, centralized decision-making with all the information of the factory might be inefficient to quickly respond to the dynamic manufacturing environments.
Therefore, multi-agent architectures with distributed decision-making have been introduced in manufacturing systems to improve flexibility and agility~\cite{leitao2009agent,shen2006agent,kovalenko2022toward}.

Some existing multi-agent architectures consist of agents who are responsible for making scheduling decisions after collecting information from PAs and RAs.
The contact agent introduced by~\cite{lepuschitz2010toward} and the rescheduling agent developed by~\cite{uhlmann2022hybrid} receive resource disruption information from the disrupted RA and then start the rescheduling process with knowledge of the entire system.
However, these types of agents essentially provide centralized decision-making for scheduling, which has limitations in quickly responding to dynamic environments.
Therefore, this section focuses on the distributed decision-making process via agent coordination to solve a rescheduling problem.

In existing multi-agent architectures,
it is commonly stated that RAs are the class of agents that identify resource disruptions by continuously collecting data from their associated resources, while the decision-making process for rescheduling could be triggered by a PA or an RA.
In the studies~\cite{wong2006integrated, kovalenko2019dynamic, antzoulatos2017multi}, PA is used to trigger the rescheduling process and determine a new schedule.
Once the disrupted RA informs the PA of a need for rescheduling, the PA sends a rescheduling request and triggers the PA-RA coordination to generate a new resource allocation schedule based on the remaining tasks and resource capacities.
In~\cite{kovalenko2019dynamic}, the RAs receiving the PA's request will propagate the request to other RAs if they cannot satisfy the requirements.
However, these methods do not try to preserve the initial schedule, thus they have a high probability of causing deviations between the new and initial schedule. 
In the rescheduling problem, the deviation between the new and initial schedule is defined as scheduling robustness~\cite{vieira2003rescheduling}. 
Thus, the methods from~\cite{lepuschitz2010toward, wong2006integrated, kovalenko2019dynamic} have limited scheduling robustness.

To address the limitation, determining modifications to the initial schedule should be considered. 
Therefore, for resource disruptions, RA coordination can be applied by using the local view of the RAs.
\cite{farid2015axiomatic} introduces a reconfiguration agent as a mediator for RA coordination to respond to the reconfiguration and communication requests from different RAs.
For direct RA coordination, some collaborative mechanisms are provided to enable a disrupted RA to request all of the other RAs~\cite{rodrigues2018decentralized} or all RAs of the same type~\cite{park2012autonomous} to make reallocation decisions.
However, it is not necessary for the disrupted RA to communicate with all RAs since some RAs do not have the required capabilities to perform the affected operations of the disrupted RA.
Therefore, these methods create a significant communication load that will limit the agility of the system in response to a disruption.
To reduce agent communication, clustering approaches have been used to provide a structured coordination process. 
In~\cite{maturana1999metamorph}, an RA cluster is defined as a set of RAs that collaborate to complete a sub-task.
\cite{barata2003coalitions} defines an RA cluster based on both physical constraints and resource proximity.
However, for a rescheduling problem, fixed coordination rules and considering only nearby resources might cause resource overload or for no alternative resource to be found, which reduces throughput and resource utilization.

To cope with the problem, the disrupted RA needs to dynamically determine the agents it coordinates with (i.e. RA cluster) instead of following a pre-defined rule-based coordination strategy since the rescheduling scenarios are highly variable.
The environment information, such as other agent attributes, and coordination behavior should be designed and included in the RAs' knowledge base.
The existing studies~\cite{farid2015axiomatic,lepuschitz2010toward, rehberger2016agent, park2012autonomous}, focus on how an RA makes decisions to respond to other agents' requests through their proposed modularized RA architectures.
\cite{kim2020multi} introduces a reinforcement learning approach to enable agents to learn the environment to solve the scheduling problem.
However, these methods do not cover how the agents can dynamically determine their coordination behaviors for the rescheduling problem.

Uncertainties and risks in the manufacturing system have also been studied for the scheduling/rescheduling problem recently.
These risk assessment methods focus on robust scheduling, which refers to deriving schedules that are resilient to disruptions~\cite{vieira2003rescheduling}.
In~\cite{mejia2020robust, lee2013risk, anbarani2022risk}, risk scenarios are incorporated into the Petri Net model or automata of the entire system and are considered when the system generates a production schedule.
\cite{liu2020parallel} provides an algorithm for robust scheduling considering uncertain processing times.
\cite{klober2017predictive} introduces a conceptual structure that enables risk assessment in production scheduling.
These studies primarily focus on risk assessment in the process of generating an initial schedule and obtaining a resilient schedule in the presence of disruptions.
However, there are disruptions that will require the development of a new schedule in order to meet the process throughput.
Therefore, incorporating risk assessment into the rescheduling process is an important need.
However, the current studies use centralized methods to cope with risks for the rescheduling problem~\cite{guo2021sequencing, framinan2019using}, while none of the existing distributed rescheduling methods incorporate risks in their decision-making process.


In summary, for the rescheduling problem, existing multi-agent decision-making methods do not currently satisfy these following needs defined to achieve agile and robust rescheduling: (1) minimization of changes to the original production schedule, (2) dynamic and distributed decision-making via agent coordination, and (3) incorporation of metrics that quantify risks into distributed rescheduling decision-making.

\section{Problem formulation}
\label{sec:problemformulation}

In this section, formal definitions of the multi-agent architecture and components within a production schedule are provided. A resource reallocation problem in the form of a rescheduling task is also formulated.
\subsection{Definitions}
\label{sec:definition}

\textbf{Manufacturing system} -- resources that are connected by material and information flow with a control architecture to produce finished goods~\cite{kovalenko2022toward}.

\textbf{Resources} -- the entities, such as humans or equipment, that perform operations (e.g., production, maintenance, and transportation) in a manufacturing system.

\textbf{Central knowledge base} -- contains all the information relevant to the manufacturing system, such as product requirements, resource capabilities, etc. 
It is initialized by the manufacturer.

\textbf{Production goal} -- an objective to transform raw materials into finished products to meet customer demands through certain resource operations. 

\textbf{Production schedule} -- a plan that specifies  resources to perform operations on parts at certain times to achieve the production goal.
A detailed definition is stated in Section~\ref{sec:productionschedule}.


\subsection{Agent formulation}
\label{sec:agentformulation}

In this work, product agents (PAs) and resource agents (RAs) are used to describe the multi-agent manufacturing system and outline the rescheduling problem.

\subsubsection{Product agent}

A PA is responsible for fulfilling the desired production requirements of its associated physical product. 
\cite{kovalenko2019model} introduced a model-based PA architecture that enables PAs to make intelligent decisions to guide products and track the production progression through the manufacturing system.
Each PA stores the status of its associated product as a discrete state in the set $X=\{x_0,x_1,...,x_f\}$, 
where $x_0$ is the initial state and $x_f$ is the final state of the product in the manufacturing system.
Each state is comprised of two elements, $x_i=(x_i^{\ell},x_i^c)$, where $x_i^\ell$ and $x_i^c$ denote the product's location and physical composition, respectively. 
Note that precedence constraints may exist in the physical composition states $x^c_i$ while usually not in the location states $x^{\ell}_i$.
For instance, a PA state can be represented as: $x_i=($ ``at machine1'', ``with a milled pocket''$)$. 

\subsubsection{Resource agent}

An RA provides high-level control for a physical resource to perform operations on products. 
In this work, RAs are grouped into two RA classes: transportation and transformation RAs, based on the operations they can perform on the products.
The resource operations are modeled as a set of discrete events, denoted by $E=\{e_0,e_1,...,e_n\}$. 
An event for a transportation RA results in a state change in the location of a product, while an event for a transformation RA results in a change in the physical composition.
More information regarding the resource agent is discussed in Section~\ref{sec:RAarchitecture}.

\subsection{Production schedule}
\label{sec:productionschedule}
From Section~\ref{sec:definition}, 
the production schedule for the manufacturing system is a collection of schedules for all $p$ products (or equivalently all $r$ resources) in the manufacturing system.
The production schedule for a PA or an RA contains different information.
The set of all PAs and RAs in the system is denoted by $A=\{PA_0,...,PA_p,RA_0,...,RA_r\}$, where $p$ is the number of PAs and $r$ is the number of RAs in the system.
The production schedule for each agent is calculated by a function:

$S: A\rightarrow (s, Ag, T_s)$, where

\hspace*{0.8em}$s:A\rightarrow e_0, ..., e_{a}:$ is a function that maps agents to the sequence of events scheduled to be performed either on that product or by that resource

\hspace*{0.8em}$Ag:s\times A \rightarrow A:$ is a function that represents the relationship that describes the events, the RAs that perform the events, and the PAs on which the events are performed

\hspace*{0.8em}$T_{s}:s(A)\rightarrow (\mathbb{R}_+,\mathbb{R}_+):$ is a function that maps events to start and end times

For a given resource, the event sequence $s(RA_j)$ represents the events that $RA_j$ will perform, and $Ag$ provides the PA on which the events are performed.
$T_s(s(RA_j))$ provides the start and end times for each event in the resource schedule. 
It is assumed that a resource cannot perform multiple events at the same time, thus there should be no intersections (and generally should exist time gaps) between event time periods for a given resource. Therefore, one might define an idle time interval between the end time of one event and the start time of the next event.
The set of idle time intervals, denoted by  $I=\{[t_0,t_1],[t_2,t_3],...\}$, can be calculated for each (re)scheduling purpose.
The production goal will be achieved if the specified resources follow their designated schedules for each product in the system.

For a specific product, the event sequence $s(PA_i)$ defines the PA state transitions from the initial state, $x_0$, to the final state, $x_f$, as stated in Section~\ref{sec:agentformulation}.
To represent how an event represents a change in the state of a $PA_i$, a state transition function is defined as $Tr:X\times E\rightarrow X$.
The PA states and event sequence, $s(PA_i)$, satisfy the transition relationship: 
\begin{equation}
x_f = T_r(x_0,s(PA_i)).
\label{eq:parttransition}
\end{equation}
Based on the transition relationship, the start and end times in $T_{s}(s(PA_i))$ indicate the time periods during which the product is associated with a specific state.
Since the product state is always defined, the times provided in $T_{s}(s(PA_i))$ will not contain any time gaps, hence, the end time of one event (or state) equals the start time of the next event (or state).
Note that an event type (e.g. milling a pocket) can occur multiple times in $s(PA_i)$, but at varying occurrence times and with different RAs.
The function $Ag$ identifies the specific RA that is associated with a particular event being applied to $PA_i$.
Note that every event in the schedule of a given product is also an event for the associated resource and vice versa. However, the indices of the specific event are not the same within the product and resource schedules.

\subsection{Problem statement}
\label{sec:problemstatement}
\begin{figure}[!t]
\centering
\includegraphics[width=1\columnwidth]{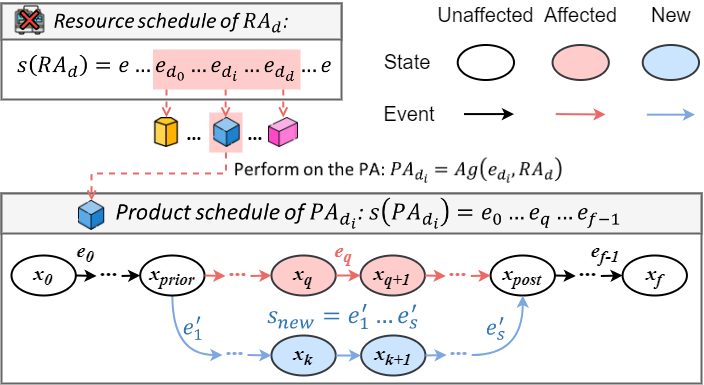}
\caption{Discrete event system representation of the problem formulation. Each affected event $e_{d_i}$ from the affected resource is associated with a specific PA ($PA_{d_i}$), where the PA schedule denotes the index as $q$. $e_q$ cannot be performed on the product and thus the state transition for $PA_{d_i}$ from $x_{prior}$ to $x_{post}$ cannot be achieved. The reallocation problem is to find a new event sequence $s_{new}$ that can recover this transition.}
\label{fig:problemformulation}
\end{figure}

Resource allocation can be formulated as a production scheduling problem~\cite{shen2006agent}.
When unexpected resource disruptions (e.g. breakdowns) occur in dynamic manufacturing systems, the initial production schedule cannot be executed as originally planned~\cite{zhang2017flexible}.
Therefore, the products that are impacted by this disruption may be rescheduled through the reallocation of resources~\cite{abumaizar1997rescheduling}.

This rescheduling problem is outlined as: given a manufacturing system ($r$ resources) with a production goal ($p$ products to produce) and feasible initial production schedule ($S$), assuming a single resource breaks down ($RA_d$), find a new feasible and resilient production schedule on-the-fly that minimizes changes to the initial schedule $S$ and optimizes user-defined objectives.
To formulate the problem, the following assumptions are provided: 
\begin{enumerate}[label={A.\arabic*}]
    \item The initial production schedule is predetermined and will achieve the production goal if followed. \label{a1}
    \item Unexpected resource disruptions are detectable by the associated RAs and result in the specific resources becoming unavailable for a certain amount of time. \label{a2}
    \item The manufacturing system contains resource redundancy and is operating with available capacity. \label{a3}
    \item The rescheduling time can be neglected compared to operation time. \label{a4}
\end{enumerate}

\ref{a1} ensures that the manufacturing goal can be met if the rescheduling process follows the production requirements in the initial schedule. 
\ref{a2} guarantees that a disruption will be identified by a resource if it occurs and also designates how a resource will be impacted by the disruption.
\ref{a3} is necessary to enable agent coordination and part rerouting.
\ref{a4} simplifies the rescheduling problem by assuming there are no changes in the manufacturing system during the decision-making process.


Once a resource disruption occurs, the associated RA is able to identify the disruption (\ref{a2}) and determine the events that the resource can no longer perform, denoted by $E_d$, which is a sub-sequence\footnote{For simplicity, the symbol $\subseteq_{seq}$ is used to represent the sub-sequence relationship in this work.} of the original event sequence for resource $RA_d$: $E_d\subseteq_{seq} s(RA_d)$.
All of the events in $E_d$ need to be re-assigned to alternative resources, which requires resource redundancy and available capacity (\ref{a3}).

\begin{algorithm}[t]
    \caption{Identify the shortest event sequence that needs to be replaced}
    \label{alg:findsd}
    \begin{algorithmic}[1]
    \REQUIRE $e_{d_i},RA_d,s,Ag$
    \ENSURE  $s_d$\\
    \COMMENT{\ \textit{Identify the index of $e_{d_i}$ in its associated product schedule}}
    \STATE $PA_{d_i}\gets Ag(e_{d_i},RA_d)$
    \STATE $j\gets 0$ and $e_j\in s(PA_k)$
    \WHILE{$e_j\neq e_{d_i}$\ \text{or}\ $Ag(e_j,PA_k)\neq RA_d$}
    \STATE $j\gets j+1$ 
    \STATE $q\gets j$
    \ENDWHILE\\
    \COMMENT{\ \textit{Find the event sequence that needs to be replaced}}
    \STATE Add $e_q$ to $s_d$\\
    \COMMENT{\ \textit{For the events $e_j$ before $e_q$ in $s(PA_k)$}}
    \STATE $j\gets q-1$
    \WHILE{$x_{j+1}^\ell=x_d^\ell$ and $j\geq0$}
    \STATE Add $e_j$ to the first position of $s_d$
    \STATE $j\gets j-1$
    \ENDWHILE\\
    \COMMENT{\ \textit{For the events $e_j$ after $e_q$ in $s(PA_k)$}}
    \STATE $j\gets q+1$
    \WHILE{$x_{j+1}^\ell=x_d^\ell$ and $j\leq f-1$}
    \STATE Add $e_j$ to the last position of $s_d$
    \STATE $j\gets j+1$
    \ENDWHILE
    \RETURN $s_d$
    \end{algorithmic}
\end{algorithm}

As shown in Fig.~\ref{fig:problemformulation}, each event $e_{d_i}\in E_d$ belongs to the schedule of its associated PA, denoted by $PA_{d_i}=Ag(e_{d_i},RA_d)$.
Since $RA_d$ cannot perform $e_{d_i}$, $PA_{d_i}$ cannot achieve its production goal (i.e., state transitions in Eqn.~\ref{eq:parttransition}).
The sequential events associated with $e_{d_i}$ in a given product schedule may become unnecessary (e.g., transportation events to/from the broken machine).
We define $s_d$ as the shortest sequence that contains $e_{d_i}$ and should be replaced by a new event sequence $s_{new}$ in the production schedule.
To identify the sequence $s_d$, the index of $e_{d_i}$ for the specific product, $PA_{d_i}$, is denoted by $q$ (i.e. $e_{d_i}=e_q$).
Algorithm~\ref{alg:findsd} determines $s_d$ by checking whether the associated states of the sequential events are related to $RA_d$.
In this way, $s_d$ is guaranteed as the shortest sequence that contains $e_{d_i}$ and needs to be replaced.
Once $s_d$ is identified, two states $x_{prior}$ and $x_{post}$ are defined as the states before and after $s_d$ in the product schedule of $PA_{d_i}$, where $Tr(x_{prior},s_d)=x_{post}$.

Therefore, for each affected event $e_{d_i}$ and its associated product $PA_{d_i}$, the rescheduling process aims to search for a new sub-sequence of events ($s_{new}$) that includes the events that need to be replaced, $s_d$:
\begin{equation}
    Tr(x_{prior},s_{new})=x_{post}
\label{eq:goal}
\end{equation}
Note that the affected event $e_{d_i}$ being performed by an alternative RA should be an element of $s_{new}$ and the new sequence should satisfy the production requirements.

Through Algorithm~\ref{alg:findsd} and Eqn.~\ref{eq:goal}, the rescheduling problem is formulated in a way that minimizes the changes to the initial schedule.
Therefore, instead of resolving a system model to generate a fully new optimal schedule (e.g., job shop schedule, which is NP-hard), we focus on modifying the initial schedule by locally searching for an alternative resource to replace the broken resource to recover the performance and thus minimize the impact to the initial schedule.
It requires less computational effort than re-generating the total schedule while it loses some optimality.
In this case, the problem in this work is polynomial-time solvable since the worst case is to evaluate all the resources in the system for each event that needs to be replaced.
This process takes $O(r\times\sum_{e_{d_i}\in E_d}|s_{d_i}|)$ computations, where $r$ is the number of RAs in the system.
A centralized method can be applied to provide an optimal schedule based on the performance objective (e.g., throughput); however, this approach requires significant computation efforts to achieve centralized optimization, making it less agile in many disruption scenarios. 
In this work, we propose an RA communication strategy with capability heuristics to avoid the need to communicate and optimize across all of the resources, thus reducing computational efforts.
We then incorporate risk assessment into the rescheduling decision-making problem to investigate how the consideration of risk improves throughput.

\section{Resource agent architecture}
\label{sec:RAarchitecture}

\begin{figure*}[!t]
\centering
\includegraphics[width=1.8\columnwidth]{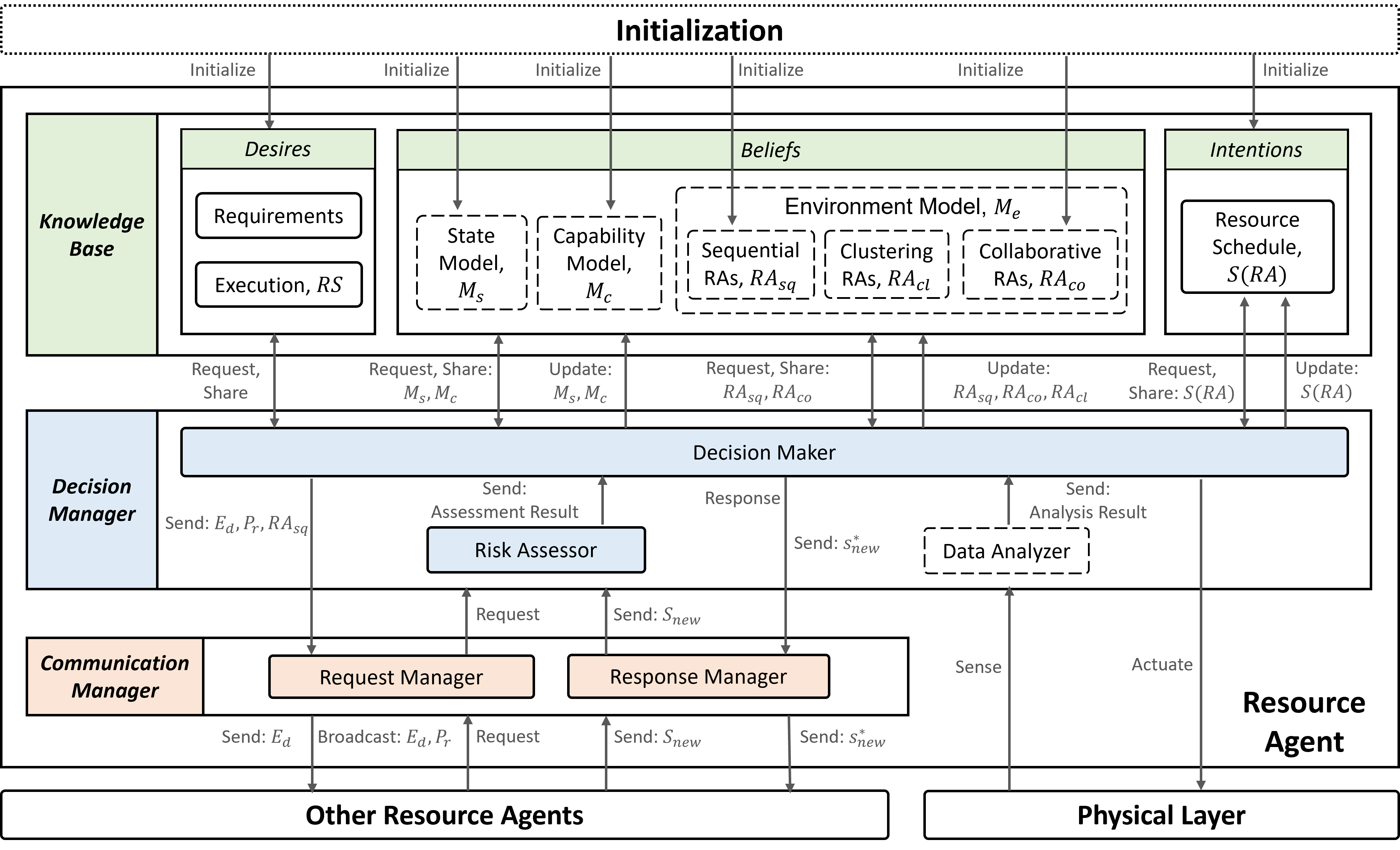}
\caption{The internal resource agent architecture, including Knowledge Base, Communication Manager, and Decision Manager. The communication between each component for the rescheduling problem is also displayed. Note that the modules with dash border are from existing studies~\cite{bi2021dynamic,toothman2021trend}}
\label{fig:RAarchitecture}
\end{figure*}

The proposed RA architecture consists of three components: a Decision Manager, a Communication Manager, and a Knowledge Base. 
A detailed design of the proposed RA architecture, including specific components and component-to-component information exchange, is shown in Fig~\ref{fig:RAarchitecture}. 
Note that this section describes the RA architecture from the perspective of a single RA, denoted by $RA_j$, in a multi-agent manufacturing system.

\subsection{Decision manager}
\label{sec:decisionmanager}
The Decision Manager is responsible for the deliberation and reasoning process of an RA.
Different decisions, such as product scheduling~\cite{kovalenko2019model}, RA response~\cite{rehberger2016agent}, etc., have been introduced in the literature.
The Decision Manager in this paper makes decisions about data analysis, scheduling management, communication, and risk assessment.

\textbf{Data analyzer} -- a component that collects and analyzes data from the physical resource through sensors.
The data analyzer may contain different data-driven models to abstract information that can be used by the agents from the raw data obtained from various sensors.


\textbf{Risk assessor} -- a component that provides enhanced deliberation and reasoning processes to the RA by evaluating the risk of decision candidates.
The risk assessor may contain different function models to analyze the risk of any decision based on the current status of the agent and the responses received from other agents.

\textbf{Decision maker} -- a component that makes decisions regarding the execution of the current schedule and responds to requests from other agents based on the current status of the RA.

\subsection{Communication manager}
\label{sec:commnucationmanager}
The Communication Manager of an RA provides the interface between the RA and other agents for exchanging information.
While the communication component has been mentioned in~\cite{kovalenko2019model,lepuschitz2010toward,park2012autonomous}, these methods do not specify the different types of communication between the RAs.
The Communication Manager in this paper includes a request manager and a response manager.

\textbf{Request Manager} -- a component that sends requests from the decision manager to other agents and passes requests received from other agents to the decision maker.

\textbf{Response Manager} -- a component that sends the response from the decision manager to other agents and passes responses received from other agents to the decision maker.

\subsection{Knowledge base}
\label{sec:knowledgebase}
The belief-desire-intention (BDI) architecture has been widely used to provide a modular framework to design intelligent agents~\cite{howden2001jack}.
Following the BDI design, the model of an RA in the authors' previous work is contained in the beliefs segment of the architecture within this work.
In this work, we reformulate the structure and content of the belief section of the RAs as the desires and intentions are developed and integrated into the Knowledge Base. 
As shown in Fig.~\ref{fig:RAarchitecture}, several aspects of the Knowledge Base are initialized before the manufacturing system begins operating. We assume this initialization is completed by the manufacturers based on the customer order, physical layer, and initial production schedule.


\subsubsection{Intentions and Desires}

\paragraph{Intentions}
Represent the plan an agent has committed to execute.
In this paper, the intentions of an RA are represented by the resource schedule $S(RA_j) = (s, Ag, T_{s})$, as defined in Section~\ref{sec:productionschedule}.

\paragraph{Desires}
Represent the goal and requirements for an agent.
As shown in Fig.~\ref{fig:RAarchitecture}, the desire of an RA is to execute the resource schedule $S(RA_j)$ without violating requirements for production and safety.

In this paper, function $P_q:E\times PA\rightarrow Requirements$ maps each scheduled event for a given product agent to its specific production requirements (e.g. precision).
The production requirements that $RA_j$ must satisfy based on the products that will engage with the given RA are represented as a set $\{P_q(e_i,Ag(e_i,RA_j)):  e_i\in s(RA_j)\}$.
These requirements are then split into hard and soft requirements.
Hard requirements must be followed while soft requirements can be negotiated to meet the demand with the introduction of a penalty.
For example, a hard requirement might be the size constraint of a product that can be assigned to a resource such that the product will fit within the workspace of the resource. A soft requirement could include the bound on the energy cost for a given event that may need to be violated in order to meet the product due date~\cite{kovalenko2022cooperative}.

The intentions and desires are related to the resource schedule, thus they are assigned to $RA_j$ once the initial production schedule of the manufacturing system is determined, and will be updated as the resource schedule changes.

\subsubsection{Beliefs}
Building from the architecture used in previous work~\cite{bi2021dynamic}, the beliefs of an agent are comprised of the state, capability, and environment models.
These models are dynamically updated (i.e., extended, shrunk, and revised) as the resource and its environments change.

\paragraph{State model}
Describes how an RA monitors the status of the associated physical resource.
\cite{qamsane2019dynamic} introduced a finite state machine (FSM) framework to model the status of a manufacturing resource using several states and transitions. 
Similarly, the RA state model in this paper is defined as an FSM that includes \textit{Idle}, \textit{Up}, and \textit{Down} states as well as the transitions between these states.

Transitions between RA states are triggered by the decision maker of the RA.
As shown in Fig.~\ref{fig:RAarchitecture}, the sensor data in the physical layer is collected by the data analyzer, which utilizes this data to identify the current status of the physical resource.
Though this paper does not focus on data-driven analysis, related work has been done to achieve state and anomaly identification~\cite{toothman2021trend}.
Having obtained the analysis results, the decision maker checks the current state model and decides whether an update to the state model is needed (e.g. trigger the transition to \textit{Down} if the resource is broken).

\paragraph{Capability model}
\label{sec:Mc}
Provides a detailed description of the operations that a resource can perform on parts.
As defined in Section~\ref{sec:agentformulation}, RAs are grouped into two RA classes: transformation RAs and transportation RAs.
The resource operations can be modeled as discrete events that drive state changes in the parts. 
Therefore, an FSM can be used to model the capabilities of an RA~\cite{kovalenko2019dynamic,balta2021model}:

$M_c = (X, E, T_r, T, A_t)$:\\
\hspace*{1.5em}$X=\{x_0, ..., x_n\}:$ a set of states that can be achieved on \hspace*{1.5em} products utilizing the resource\\
\hspace*{1.5em}$E=\{e_0, ..., e_m\}:$ a set of events representing operations \hspace*{1.5em} that change product states\\
\hspace*{1.5em}$T_r:X\times E\rightarrow X:$ a state transition function\\
\hspace*{1.5em}$T: E \rightarrow \mathbb{R}_+:$ amount of time associated with an event \\
\hspace*{1.5em}$A_t(E,RA_j):$ a function that maps events and specific \hspace*{1.5em} RAs to the physical resource attributes associated with \hspace*{1.5em} each event (e.g. payload limitations)



In the capabilities model, 
the state set $X$ contains all changeable states for the products associated with the given resource.
$E$ and $T_r$ follow the definition provided in Section~\ref{sec:agentformulation}.
The transition function $T_r$ is inherited from the PAs.
$T$ represents the nominal cost (denoted as operation time) for each event to occur assuming the cost for the same event is identical for different products.
$A_t$ provides the resource attributes for each event.
Note that multiple RAs in the same class could have the same events, but the attributes might be different. 
The characteristics of the attributes are described by parameters, such as the speed limitation, payload, and part dimensionality. 

As shown in Fig.~\ref{fig:RAarchitecture}, the capability model is initialized based on the physical manufacturing system. 
Similar to the state model, as the data analyzer receives information constantly from the resource in the physical layer, the decision maker updates the capability model if there are any changes to the resource.
The changes could be manual, such as tool replacement/removal, or spontaneous, such as machine breakdown.

\paragraph{Environment model}
The RA's knowledge of other RAs in the system is captured in the environment model.
The relationships to these agents are modeled as mapping functions that map events or states to different sets of RAs, namely clustering, sequential, and collaborative RAs.\\
    \noindent \textbf{Clustering RAs:} The clustering RAs for $RA_j$ are the set of RAs that can perform the same events as $RA_j$ for a given subset of events $E_s$ in $RA_j$'s capability model.
     Each event in the given event subset $e_i\in E_s$ corresponds to a unique cluster.
     The relationship between each event $e_i$ and the associated clustering RAs is modeled as a cluster mapping function, which is defined as $C_\ell:E_s\times RA\rightarrow 2^{RA}$, where
     \begin{equation}
         C_\ell(e_i, RA_j)=\{RA_k\ |\ e_i\in E_{RA_k},\ RA_k\neq RA_j\}
     \label{eq:cluster}
     \end{equation}
     $2^{RA}$ denotes the power sets of the RAs in the manufacturing system. 
     $E_{RA_k}$ represents the event set in the capability model of $RA_k$.
     Therefore, the set of $C_\ell(e_i,RA_j)$ maps represents the clustering RAs for the given event subset $E_s$ for $RA_j$.
     As shown in Fig.~\ref{fig:RAarchitecture}, clustering RAs are not formed during the initialization.
     When clustering RAs are needed, $RA_j$ retrieves the capabilities of other RAs from the centralized knowledge base and checks the constraints in Eqn.~\ref{eq:cluster} to form the cluster map. Alternatively, $RA_j$ can also request the capability information from the RAs within the manufacturing system to form the clusters dynamically.  \\
\noindent \textbf{Sequential RAs:} 
    The sequential RAs of $RA_j$ depend on the resource schedule and associated product schedules.
    Every event in $RA_j$'s schedule $e_i\in s(RA_j)$ corresponds to a specific product agent $PA_k=Ag(e_i,RA_j)$.
    In the product schedule of $PA_k$, the RAs that perform the events directly before and after $e_i$ are sequential RAs of $RA_j$ for this specific event.
    Each event $e_i\in s(RA_j)$ corresponds to a unique set of sequential RAs.
    To identify the sequential RAs, the index of event $e_i$ in the product schedule of $PA_k$ is denoted as $q$, which can be found following the same process in Algorithm~\ref{alg:findsd}.
    Note that $q$ is bounded by $0\leq q\leq f-1$ since the event sequence in the product schedule of $PA_k$ is defined as $s(PA_k)=e_0...e_{f-1}$.
    Therefore, if $e_i$ is the first or last event in the product schedule of $PA_k$, there is only one sequential RA for this event $e_i$.
    Otherwise, there are two sequential RAs.
    $RA_j$ stores the information about the sequential RAs in a map that relates the scheduled event to specific RAs: $s(RA_j)\rightarrow 2^{RA}$.
    The set of the sequential RAs depending on the index $q$ defined above:
    \begin{equation}
        \begin{split}
        \{Ag(e_{q+1},PA_k)\},\ & \text{if}\  q=0\\
        \{Ag(e_{q-1},PA_k)\},\ & \text{if}\ q=f-1\\
         \{Ag(e_{q\pm1},PA_k)\}, \ & \text{if}\ 0<q<f-1
        \end{split}
     \label{eq:sequential}
     \end{equation}
     The sets of $S_q(e_i)$ represent the sequential RAs for the scheduled events of $RA_j$.
     The set of sequential RAs is formed in $RA_j$'s Knowledge Base based on the initial production schedule.
     As the system runs, the schedule of agents might need to change to adapt to disruptions, thus the sequential RAs should be updated as the schedule changes.

   \noindent \textbf{Collaborative RAs:} To represent the collaboration between RAs,~\cite{kovalenko2019dynamic} introduced neighboring RAs, which have shared states in the capability model.
    In this work, collaborative RAs of $RA_j$ are defined as the set of RAs that contain the same location states in their capability model.
    Each location state $x^\ell_i$ in $RA_j$'s capability model corresponds to a unique set of collaborative RAs.
    This relationship is modeled as a mapping: $X\times RA\rightarrow 2^{RA}$.
    Therefore, for $RA_j$, its collaborative RAs in terms of $x^\ell_i$ is described as:
    \begin{equation}
         \{RA_k\ |\ x_i^\ell\in X_{RA_k},\ RA_k\neq RA_j\}
     \label{eq:collaborative}
     \end{equation}
     where $X_{RA_k}$ is the state set in the capability model of $RA_k$.
     Note that a transportation RA can have both transportation and transformation collaborative RAs, while a transformation RA can only have transportation collaborative RAs.
     The set of collaborative RAs is formed in $RA_j$'s Knowledge Base as the RA and its capability model are initialized, following the state relationship discussed above.
     This set of collaborative RAs is updated as the capability model changes.
     For example, if a mobile robot can no longer reach a machine at $x^\ell$, the states related to $x^\ell$ will be removed from its capability model, as well as the collaborative RAs related to $x^\ell$.


\section{Resource agent coordination for rescheduling}
\label{sec:reschedulingprocess}

In this section, the proposed rescheduling strategy via RA coordination is described.
The coordination is guided by the agent environment models instead of following pre-defined rules.
These models can be easily updated and scaled for different systems, thus the agent coordination behaviors are flexible and adaptable.
An overview of the rescheduling process is shown in Fig.~\ref{fig:flowchart}.
In the agent coordination process, the constraints of the schedule are checked when agents determine their responses, and the new event sequence is augmented by propagating requests until the state transition is satisfied.
New aspects based on the authors' previous work~\cite{bi2021dynamic} include: (1) identification and sorting of the affected events based on priority, (2) requirements relaxation in the cluster formation, and (3) risk assessment of the new schedules based on uncertainties.

\subsection{Rescheduling request}
\label{sec:reschedulingrequest}
When a resource breaks down, the associated RA, denoted by $RA_d$, must identify the breakdown, determine events that are affected by the breakdown, and create bid requests to start the rescheduling process.

\subsubsection{Identify disruption and affected events}
\label{sec:identifydisruption}
A resource agent collects data continuously from the associated physical resource through sensors attached to this resource.
The data is passed into the data analyzer within the Decision Manager of the RA.
By feeding the data to the models in the data analyzer, the data analyzer identifies the current status of the physical resource and sends this information to the decision maker.
The decision maker then updates the knowledge base of the RA.
When a resource is broken, the RA identifies the disruption and updates the state model to indicate a \textit{Down} state.

After identifying the breakdown, the decision maker requests information about the resource schedule $S(RA_d)$ and production requirements $P_q$ from the Knowledge Base.
The decision maker will then determine the sequence of events, denoted as $E_d$, that need to be rescheduled.
$E_d=e_{d_0}e_{d_1}...e_{d_d}$ is a priority event sequence, where each event corresponds to a priority value, which is calculated based on the original start time and the priority/importance of the associated product.
An example priority mapping function could include a weighted sum of the inverse of the original start time and due date. 
For this example function, the order of the affected events in the sequence $E_d$ will increase as the start times and/or due dates of a given product are extended. 

To realize dynamic rescheduling on the fly, the proposed method reschedules the affected events in a sequential manner following the order provided in $E_d$. 
For each affected event in $E_d$, $RA_d$ runs Algorithm~\ref{alg:findsd} to identify $x_{prior}$ and $x_{post}$.
For simplicity, the following description focuses on the rescheduling process for a single affected event $e_{d_i}$. This process will be repeated for each additional event within $E_d$.

\subsubsection{Broadcast rescheduling request}
\label{sec:broadcastrequest}

\begin{figure}[!t]
\centering
\includegraphics[width=1\columnwidth]{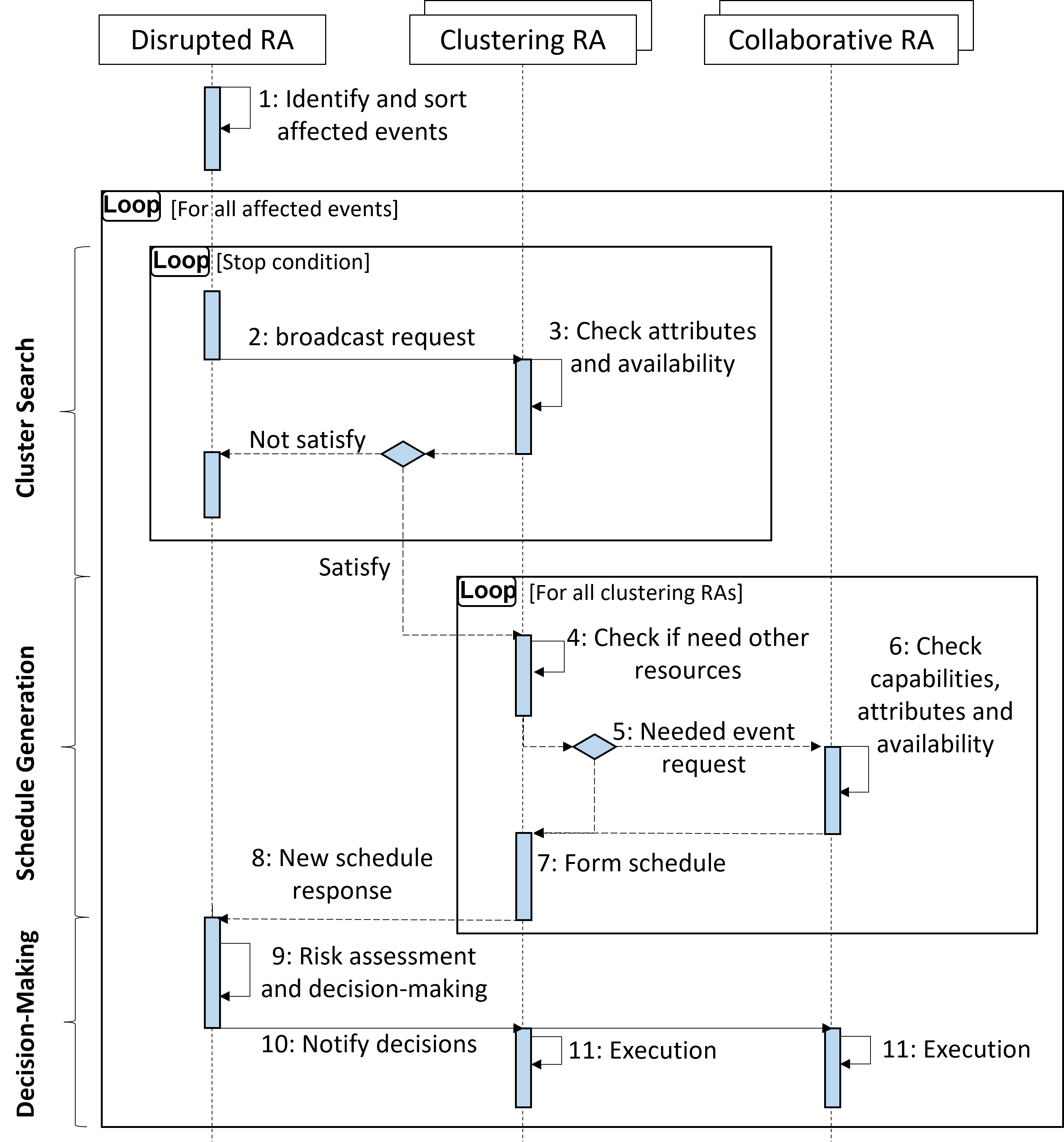}
\caption{Coordination behaviors of resource agents for rescheduling process} 
\label{fig:flowchart}
\end{figure}

After identifying and sorting the affected events, the scheduling manager of $RA_d$ sends a rescheduling request to the request manager.
A broadcast technique is used for RAs to communicate information~\cite{wooldridge2009introduction}.
For each $e_{d_i}$, the request manager can dynamically identify the cluster RAs associated with $e_{d_i}$ via the environment model and broadcasts the rescheduling bid request $Req=(e_{d_i}, P_q, x_{prior}, x_{post})$. 
$P_q$ is the function that maps $e_{d_i}$ to the production requirements.
$x_{prior}$ and $x_{post}$ define the states that denote where a transition must be rebuilt, as defined in Eqn.~\ref{eq:goal}.

As mentioned in Section~\ref{sec:problemstatement}, the sequential events of $e_{d_i}$ may become unnecessary or transition to a different sequential event depending on the RA used to replace the affected event. In the case of a change to the sequential events, the request manager of $RA_d$ sends the event $e_{d_i}$ to the sequential RAs in $S_q(e_d)$ and requests them to remove the sequential events for $e_{d_i}$ from their schedule.

\subsection{Resource agent coordination}
\label{sec:racoord}

\subsubsection{Cluster search}
\label{sec:clustersearch}

Based on the resource capabilities, only the clustering RAs of $RA_d$ with respect to $e_{d_i}$ have access to the broadcast request. 
These RAs access the rescheduling request and their request managers send the request to their decision maker.
Through this clustering scheme, the agent coordination is more effective since $RA_d$ requests the agents that can perform $e_{d_i}$ instead of requesting all the other RAs or only nearby RAs.
The decision maker requests the capability model from the Knowledge Base and conducts the following match-making steps:
\begin{itemize}
    \item Check whether the RA still contains the affected event $e_{d_i}$ in its capability model
    \item Determine whether the RA's associated resource attributes can satisfy the production requirements $P_q(e_{d_i})$
\end{itemize}
As defined in Eqn.~\ref{eq:cluster}, the first step ensures that a specific RA should be considered as a clustering RA of $RA_d$, while the second check determines whether the resource can successfully meet the production requirements (e.g. meet the hard and soft constraints). 
The hard requirements must be satisfied, while the soft requirements can be negotiated (e.g. relaxed) within a tolerance range with penalties.
A smaller, more focused clustering RA set is generated by the second match-making step:
\begin{equation}
\Tilde{P_q}(e_{d_i},Ag(e_{d_i},RA_d))\subseteq A_t(e_{d_i},C_\ell(e_{d_i},RA_d))
\label{eq:attributes}
\end{equation}
where $A_t(e_{d_i},C_\ell(e_{d_i},RA_d))$ represents the set of resource attributes of $e_{d_i}$ in the capability models of the clustering RAs.
$\Tilde{P_q}$ defines the production requirements with relaxed soft requirements.
As such, the RAs that satisfy Eqn.~\ref{eq:attributes} form a new cluster for $RA_d$ with respect to the affected event $e_{d_i}$:
\begin{equation}
\begin{split}
    \Tilde{C_\ell}(e_{d_i},RA_d)&=\{RA_c\ |\ RA_c\in C_\ell(e_{d_i},RA_d),\\ &\Tilde{P_q}(e_{d_i},Ag(e_{d_i},RA_d))\subseteq A_t(e_{d_i},RA_c)\}
\end{split}
\label{eq:newclusterset}
\end{equation}

The RAs in the cluster $\Tilde{C_\ell}(e_{d_i},RA_d)$ represent the subset of RAs that can perform $e_{d_i}$ and satisfy the production requirements of $e_{d_i}$.

\subsubsection{Schedule generation}
\label{sec:schedulegeneration}

Once the cluster $\Tilde{C_\ell}(e_{d_i},RA_d)$ is formed, each RA in the cluster follows the same process to generate a new schedule.
For simplicity, the following description focuses on one $RA_c$ in the cluster $\Tilde{C_\ell}(e_{d_i},RA_d)$.
As mentioned in Eqn.~\ref{eq:goal}, a new event sequence, $s_{new}$, needs to be formed to achieve the transitions from $x_{prior}$ to $x_{post}$ in terms of location and physical composition.
However, as defined in Section~\ref{sec:knowledgebase}, an event can only achieve either a location or physical composition transition. Therefore, the clustering RAs must verify whether the event $e_{d_i}$ satisfies the production needs given in Eqn.~\ref{eq:goal} or if other events will be needed. 

The RAs are grouped into two classes in Section~\ref{sec:agentformulation}.
If $RA_d$ is a transportation RA, then $e_{d_i}$ must be an event that drives a location change of the product and does not change the physical composition (i.e. $x_{prior}^c=x_{post}^c$).
In this case, Eqn.~\ref{eq:goal} is rewritten as:
\begin{equation}
    Tr(x_{prior},e_{d_i})=x_{post},\ \text{with}\ x_{prior}^c=x_{post}^c
\label{eq:transportation}
\end{equation}
For location events, a single clustering RA can generally replace $RA_d$ without the need for further RA coordination to form a feasible new schedule. In this example, the new schedule $s_{new}$ only contains $e_{d_i}$.

However, in the case where $RA_d$ is a transformation RA, the sequential events associated with $e_{d_i}$ that provide location transitions must be reassigned. Therefore, simply replacing $RA_d$ with a clustering RA that performs $e_{d_i}$ will not fulfill the required transitions in Eqn.~\ref{eq:goal} and other events must be included in $s_{new}$.
Using Eqn.~\ref{eq:cluster}, the clustering RA can only drive a change in the physical composition by performing event $e_{d_i}$:
\begin{equation}
    Tr(x_{prior}^c,e_{d_i})=x_{post}^c
\label{eq:transformation}
\end{equation}
To truly replace $RA_d$, the transformation clustering RA will require help from transportation RAs to move the product into and out of its location, denoted by $x_{RA_c}^\ell$.
Thus, transportation events that drive location changes from $x_{prior}^\ell$ to $x^\ell_{RA_c}$ and $x_{RA_c}^\ell$ to $x_{post}^\ell$ need to be found. 
As shown in Fig.~\ref{fig:flowchart}, the clustering RA sends requests to its collaborative RAs.
These collaborative RAs check their capability models and search for transportation events that will satisfy the location change requirements.

If the event does not exist, a propagation method can be used~\cite{kovalenko2019dynamic} to find more transportation events to drive the location change from $x_{prior}^\ell$ to $x^\ell_{RA_c}$ or $x_{RA_c}^\ell$ to $x_{post}^\ell$.
These events should be appended to $s_{new}$ to form the final schedule.

\begin{figure}[!t]
\centering
\includegraphics[width=1\columnwidth]{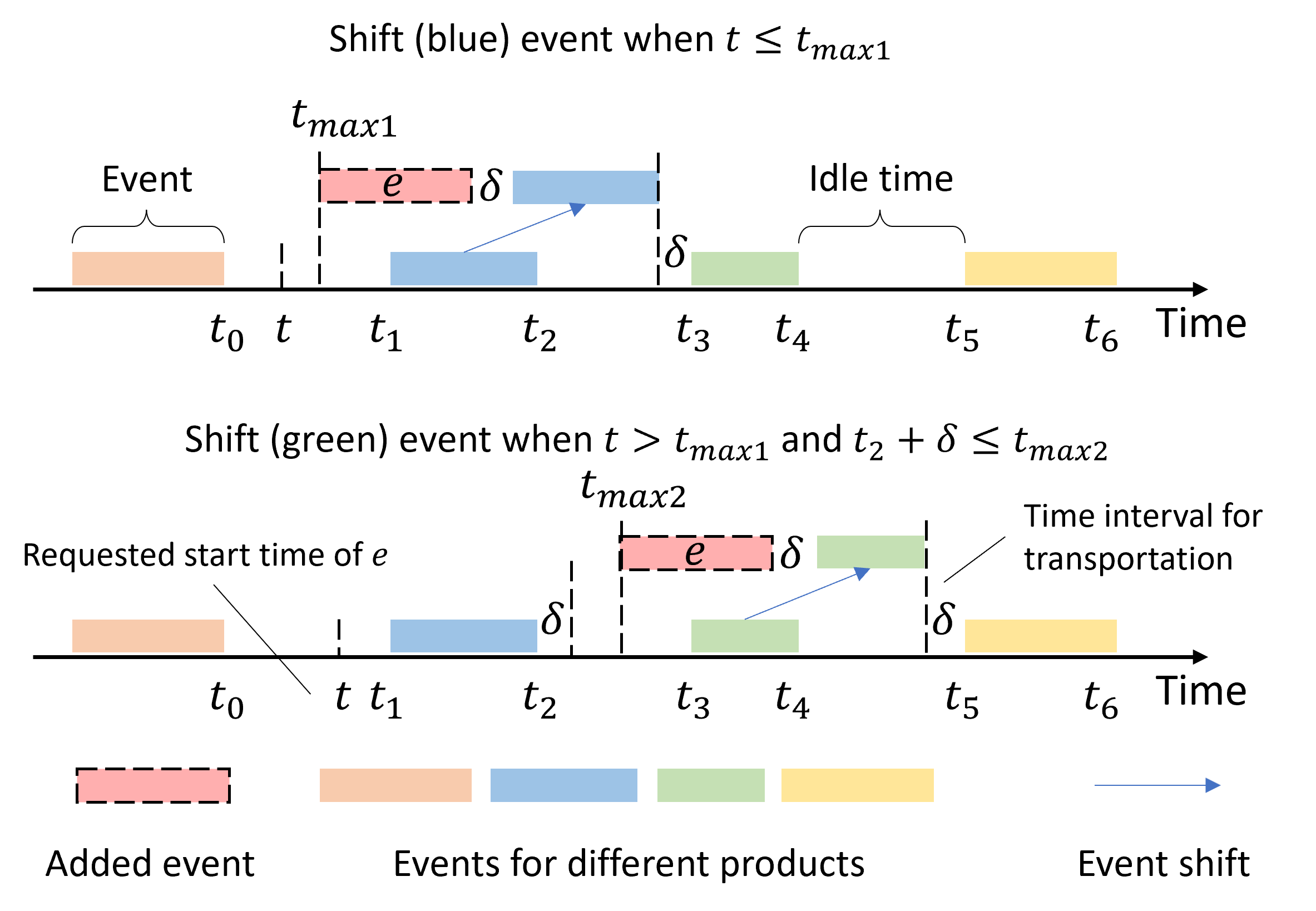}
\caption{The Gantt charts that show the schedule of a clustering RA to illustrate how the function $H$ allows one event shift} 
\label{fig:functionH}
\end{figure}

Once the event sequence $s_{new}$ is determined, the timing to perform these events needs to be determined.
In previous work~\cite{bi2021dynamic}, new events were assigned to the corresponding RA without changing the existing schedule, which led to a large delay in the product cycle time.
In this paper, a function $H$ is defined to calculate the earliest available time to start an event $e$ based on the idle time $I$ of the resource and the requested start time of $e$, denoted by $t$.
Therefore, the function $H$ serves as a heuristic that minimizes the completion time of the new event.
To simplify the time propagation, the transportation is handled by adding a time interval $\delta$ between any two operations in a machine schedule.
To minimize the effect on other events and products, only one scheduled operation is allowed to be shifted when adding $e$ to a clustering RA. This constraint determines the latest time $t_{max1}=t_3-\delta-(t_2-t_1)-\delta-T(e)$ event $e$ can be assigned to the current idle time interval, as shown in the upper Gantt chart in Fig.~\ref{fig:functionH}.
Note that $t_{max1}=\infty$ if $t_1, t_2$ or $t_3$ does not exist.
If the requested time $t$ is larger than $t_{max1}$, event $e$ cannot be scheduled before the event in $(t_1,t_2)$. If this occurs, the clustering RA will check the next idle time interval to evaluate if $t_{max2}=t_5-\delta-(t_4-t_3)-\delta-T(e)$ will provide sufficient time to assign event $e$ to this next idle time interval, see lower Gantt chart in Fig.~\ref{fig:functionH}.
Note that $t_{max2}=\infty$ if $t_5$ does not exist.
Function $H$ is defined as follows:
\begin{equation}
    H(I,t,\delta,e) =
  \begin{cases}
    t_0+\delta, & t\leq t_0+\delta\leq t_{max1}\\
    t, & t_0+\delta< t\leq t_{max1},\\
    t_2+\delta, & \max\{t,t_0+\delta\}>t_{max1}\\ &\text{and } t_2+\delta\leq t_{max2}
  \end{cases}
  \label{eq:earliesttime}
\end{equation}
where the resource idle time set, $I=\{[t_0,t_1],[t_2,t_3],...\}$, is obtained from the $T_s(s(RA_j))$ in the resource schedule, as mentioned in Sec.~\ref{sec:productionschedule}.
Note that function $H$ can be expanded if more than one scheduled operation is allowed to be shifted.

For each new event sequence $s_{new}=e_{1}e_{2}...e_{s}$, the post event should always start after the prior event ends:
\begin{equation}
    t_{s,i} + T(e_{i}) \leq t_{s,j},\ 1\leq i<j\leq s
\label{timeconstraint}
\end{equation}
where $t_{s,i}$ and $T(e_{i})$ represent the start time and time cost (e.g. cycle time) of event $e_{i}$, respectively. 
The start time of a later event $t_{s,i+1}$ is obtained from function $H$ with the requested start time of $t_{s,i} + T(e_{i})$.
Note that in the time interval $[t_{s,i} + T(e_{i}),\ t_{s,j}]$, a part remains with the current RA that performs $e_{i}$ until the RA that performs the next event $e_{j}$ is available. The combination of the clustering and collaborative RAs form a set of new event sequences, denoted by $s_{new}$, to replace $s_d$ in order to achieve the transition from $x_{prior}$ to $x_{post}$.

\subsection{Schedule risk assessment and decision-making}

\subsubsection{Risk assessment}
\label{sec:riskassessment}

When RAs send a response to form a new schedule, the information in the response may contain uncertainties.
In this work, we define uncertainty as information about a resource attribute or state that may be stochastic or probabilistic rather than deterministic. 
These uncertainties can be modeled by utilizing the manufacturing data. For example, a Gaussian distribution may be used to model uncertainty in machine operation time~\cite{liu2020parallel}. Uncertainties introduce a potentially costly effect during the decision-making process for the rescheduling problem.
We define the effects associated with variations in the rescheduling process, such as cycle time delay and schedule deviation, as risks in this paper.

To consider risks in the decision-making process, all new schedules should incorporate a risk assessment process based on the set of resources chosen to replace the affected event sequence $s_d$. 
There are two key risks considered in this work: 
\begin{itemize}
    \item $R_1$: the risk of a new event in $RA_j$ causing operational delays for the other products associated with this resource
    \item $R_2$: the risk of an added event in $RA_j$ increasing the probability of breakdown
\end{itemize}
The quantification of the two risks is discussed below through an example.
Note that the definition and quantification of risks, uncertainties, and how they are related may vary according to how a different resource may evaluate the risks. 


\paragraph{Risk of an added event causing operation delays for other products scheduled with $RA_j$}
Although event start and end times are provided in a new schedule (see Section~\ref{sec:schedulegeneration}), the actual times of these events may be shifted slightly due to uncertainties in the operation times of the events.
Note that the operations before the added event are assumed to have been completed.
As shown in Fig.~\ref{fig:functionH}, if the added event (red block with dash outline) takes longer to finish, the next event (blue block) for this resource could be impacted, which may also affect the following event (green block).
Therefore, this risk evaluates the likelihood that a posterior event will be affected.

As defined in Section~\ref{sec:schedulegeneration}, without considering uncertainties, $t_{max1}$ and $t_{max2}$ represent the latest start times for which an event can be added into the sequence without affecting a posterior event for a given resource.
The start time of the added event, $H(I,t,\delta,e)$, is obtained from Eqn.~\ref{eq:earliesttime}. 
If $H(I,t,\delta,e)$ is close to $t_{max1}$ or $t_{max2}$, the risk of causing a delay for the following event is high.
To evaluate this risk, the time deviation, denoted by $\Delta t$, between the start time and $t_{max1}$ or $t_{max2}$ is calculated:
\begin{equation}
    \Delta t =
  \begin{cases}
    t_{max1} - H(I,t,\delta,e), & H(I,t,\delta,e)=t_0+\delta\\
     & \text{or}\ H(I,t,\delta,e)=t\\
    t_{max2} - H(I,t,\delta,e),  &H(I,t,\delta,e)=t_2+\delta
  \end{cases}
  \label{eq:deltat}
\end{equation}
If we assume a Gaussian or uniform distribution for the cycle times of different events, then $t_{max1}$, $t_{max2}$, and thus $\Delta t$ are all random variables with known distributions.

Note that $\Delta t$ is non-negative, where a larger $\Delta t$ provides better tolerance to operation time uncertainty, which translates to a lower risk for causing delay.
If there are no posterior events, then $t_{max1}$ and $t_{max2}=\infty$, hence $\Delta t=\infty$, and the risk is zero.
If $\Delta t=0$, then the new event is scheduled to start at $t_{max1}$ or $t_{max2}$. Given the uncertainty in cycle times, this indicates a high risk decision.
In a new event sequence, $s_{new}=e_0e_1...e_s$, each event is added to the schedule of the specified resource.
We define the risk for a given resource $RA_j$ in the new schedule as follows:
\begin{equation}
    Q(RA_j) = 1- \mathbb{E}\left(\frac{\Delta t}{t_{max}}\right)
    \label{eq:r1}
\end{equation}
where $\mathbb{E}$ represents the expected value, and $t_{max}=t_{max1}$ or $t_{max2}$ depending on the conditions in Eqn.~\ref{eq:deltat}.
Note that as the difference between the maximum threshold and true start time increases, the risk goes down. 
Equation~\ref{eq:r1} limits the value of $Q(RA_j)$ to lie between 0 and 1.
This type of risk can be calculated for each event within the new schedule $s_{new}$.
The total value for Risk 1 associated with this schedule is defined as the maximum value among the resources that perform the new schedule:
\begin{equation}
    R_1 = \max\{Q(RA_j)\}_{j\in[0,s]}
    \label{eq:r1max}
\end{equation}

\paragraph{Risk of an added event in $RA_j$ increasing the probability of breakdown}
If a resource in the new schedule breaks down, it will lead to more rescheduling requirements for the products that have uncompleted scheduled operations by this resource.
Therefore, risk 2 is evaluated by determining the probability of breakdown if the resources in the new schedule are assigned new events to perform.
This risk is based on the assumption that every resource has a historical mean time between failure (MTBF), and that the addition of a new event will introduce more wear and tear to the resource and move the resource closer to the MTBF. 


We define the probability of resource breakdown as a function that maps RAs to a value between 0 and 1: $P_r:RA\rightarrow [0,1]$. This probability is given as:
\begin{equation}
    P_r(RA_j) = \frac{o_{c}}{o_n}
    \label{eq:prob}
\end{equation}
where $o_c$ is the number of operations that the resource has performed since the last maintenance event, and $o_n$ represents the nominal number of operations that the resource generally performs between breakdowns. 

In a new event sequence, $s_{new}=e_0e_1...e_s$, the breakdown of any resource in the new schedule makes the new schedule unsuccessful.
Therefore, Risk 2 is defined as the maximum probability of breakdown of a resource in the new schedule:
\begin{equation}
    R_2 = \max\{P_r(RA_j)\}_{j\in[0,s]}
    \label{eq:r2}
\end{equation}
Note that as the probability of resource breakdown increases, the risk goes up.
The risk value is between 0 and 1 since it is a probability calculated by Eqn.~\ref{eq:prob}.

Since a system might apply different importance levels to the different risks, the overall risk assessment value that will be incorporated into the decision-making process is a weighted sum of the risks (R1 and R2 in this work) $\sum_{i=1}^k w_i R_i(s_{new})$, where $w_i$ is the weight factor for $R_i$.

\subsubsection{Decision-making}
\label{sec:decisionmaking}

\paragraph{Determine the new schedule}
Once the risk assessment is completed, $RA_d$ is responsible for choosing a new schedule from the set of possible event sequences $S_{new}$.
Note that every event sequence in $S_{new}$ satisfies all the constraints to achieve the production goal due to the proposed problem formulation and agent coordination.
Therefore, the new schedule selection problem is reduced to obtain the new schedule that optimizes the rescheduling objectives defined by the manufacturer.
This type of optimization can be easily solved by some classical algorithms, such as bubble sort and divide-and-conquer.
Note that this optimization provides the optimal new schedule from the candidate solution set $S_{new}$.
However, the global optimal solution may not be in the candidate set $S_{new}$ since all the candidate new schedules are formed by agent local decision-making.
An example is given in Eqn.~\ref{eq:min}:
\begin{equation}
s_{new}^*=\argmin_{s_{new}\in S_{new}}\ \mathcal{J}(s_{new})
\label{eq:min}
\end{equation}
where $s_{new}^*\in S_{new}$ is the event sequence that provides minimal objective. 
The multi-objective function $\mathcal{J}$ is a sum of the cost, penalty and risk for one event sequence $s_{new}=e_{1}e_{2}...e_{s}$, as shown in Eqn.~\ref{eq:obj}:
\begin{equation}
\mathcal{J}(s_{new}) = \sum_{i=1}^s \bm{\alpha} \bm{C}(e_{i}) + \sum_{i=1}^s \beta_i p_i + W\sum_{i=1}^k w_i R_i(s_{new})
\label{eq:obj}
\end{equation}
where $C(e_i)=[C_1(e_i)\ C_2(e_i)\ \cdots\ C_n(e_i)]^T$ captures a nominal cost function for event $e_i$ based on $n$ metrics and $\alpha = [\alpha_1 \alpha_2 \cdots \alpha_n]$ describes the corresponding weights.
The pre-defined cost metrics could include operation time, finish time, energy cost, resolution, etc. 
If there are soft constraints that must be negotiated, $p_i$ denotes the penalty for performing $e_i$ and $\beta_i$ is the corresponding weight. The risks associated with the given sequence are evaluated in 
$\sum_{i=1}^k w_i R_i(s_{new})$ for a given sequence. Parameter $W$ is used to scale the risk based on the scale of the cost and penalty and what value the decision maker places on the assessment of risk.
Note that in Eqn.~\ref{eq:obj}, the objectives, penalties, and risks are defined by the manufacturers and the weight parameters depend on how the manufacturers desire to balance the objectives, penalties, and risks.
Future work will investigate the sensitivity of Eqn.~\ref{eq:obj} to changes in these values and identify a method for optimizing under various conditions.

As shown in Fig.~\ref{fig:flowchart}, the affected resource, $RA_d$, informs the RAs that will be associated with the new event sequence, $s_{new}^*$, through a Communication Manager. The new RAs receive the notification and pass the information to their Knowledge Bases to update their resource schedules and provide high-level control for their associated physical resources to perform the events.

\paragraph{No schedule found}

The result of the rescheduling problem depends on resource redundancy and available capacity, which are the assumptions we made in Section~\ref{sec:problemstatement}.
In practice, manufacturing resources are limited in a factory, therefore, the existence of a feasible new schedule is not guaranteed. 
Therefore, in the proposed method, if no schedule is found within the required constraints (e.g., no redundant resources available), the $RA_d$ will request the central controller of the manufacturing system or human manager to make further decisions or relax additional constraints.
As mentioned in Sec.~\ref{sec:problemstatement}, the centralized method evaluates all of the resources in the system for each event that needs to be replaced during the rescheduling process.
Therefore, the centralized method forms the candidate solution set $S_{new}$ by considering all the combinations of resources in the system and then solves the following optimization:
\begin{subequations}\label{mip}
\begin{align}
\label{obj:mip}
\min_{s_{new}\in S_{new}}& \hspace{1ex}\mathcal{J}(s_{new})\\
\text{s.t.}\qquad
&Tr(x_{prior},s_{new})=x_{post}\\
  \label{cst:cap_p}
& t_{s,i} + T(e_{i}) \leq t_{s,j},\ 1\leq i<j\leq s, e_i\in s_{new},
\end{align}
\end{subequations}
where these objectives and constraints are the same as those considered by the distributed method.



\section{Case study}
\label{sec:casestudy}
To evaluate the feasibility and performance of the proposed framework, the proposed RA architecture and rescheduling strategy are implemented in a simulated manufacturing system.
In this section, the set-up of the simulated manufacturing system and the results of the case study are provided.
\subsection{Case study set-up}
\begin{figure}[tb]
    \centering
    \includegraphics[width=1\columnwidth]{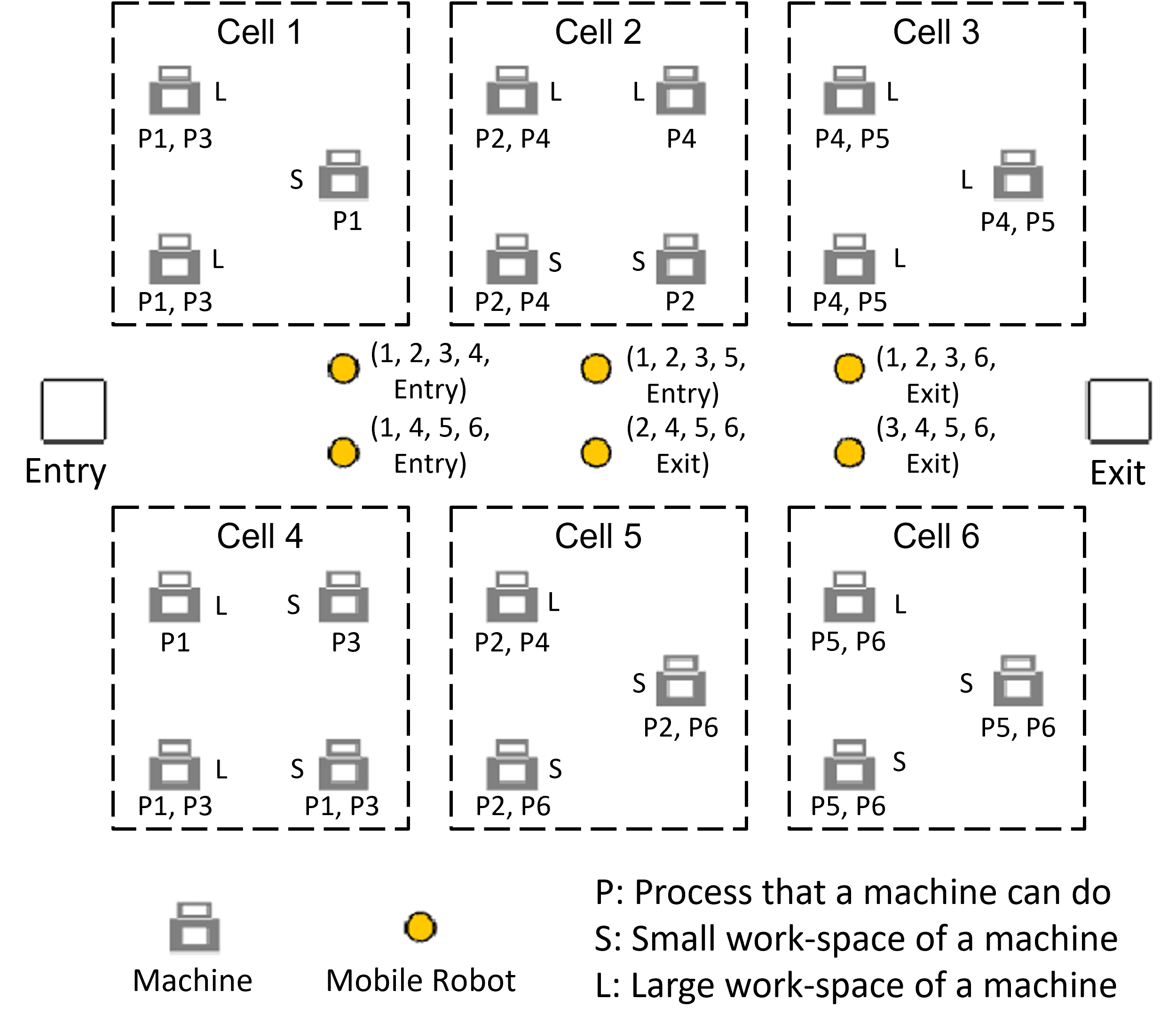}
    \caption{A screenshot of the facility layout from the RepastS environment. The annotations indicate the capability of each resource.}
    \label{fig:setup}
\end{figure}

In this study, we use a Repast Symphony (RepastS) platform~\cite{macal2006introduction} to model a multi-agent system and simulate the behavior of the agents due to its flexibility and scalability properties. 
The simulated manufacturing system represents a modified version of the Intel Mini-Fab~\cite{yoon2008multiagent}, a semiconductor manufacturing facility. 
The simulated system contains two infinite-sized buffers, Entry and Exit, and 20 machines that are connected via a network of 6 mobile robots, as shown in Fig.~\ref{fig:setup}.
The annotations represent the capabilities of the machines and mobile robots. 
There are 6 different processes (P1-P6) that the machines can perform, where the operation costs in ticks (RepastS unit of time) range from 110-200.
For example, the annotation for a machine indicates which processes it can perform and whether the workstation space is large or small. 
The annotation for a mobile robot represents which cells and buffers it can reach when moving the products.

Two types of products, labeled S (small) and L (large), are introduced into the system, where each type of product has the following process requirements:

\begin{itemize}
    \item S-product: P1 $\rightarrow$ P2 $\rightarrow$ P3 $\rightarrow$ P6
    \item L-product: P1 $\rightarrow$ P3 $\rightarrow$ P4 $\rightarrow$ P5
\end{itemize}

Machines labeled L can operate both L-products and S-products, while machines labeled S can only operate on S-products.
Products enter the system from the Entry buffer and leave the system through the Exit buffer after completing the desired processes.


\subsection{Case study and results}

In this simulated manufacturing system, 50 L-products and 50 S-products are fed alternatively into the system with a pre-generated initial production schedule.
Products enter the facility every 30 ticks starting at tick 10.
To provide an opportunity for a rescheduling event to occur, the initial production schedule is designed with 50\% resource utilization.
Uncertainty in machine operation time and the probability of machine breakdown are added to all machines in the simulated system.
The system starts operations with the probability of machine breakdown ranging from 3.3\% to 10\%.
If a machine undergoes a breakdown, a rescheduling process will be triggered.
The mean time to repair ranges from 1000-1500 ticks for a broken machine.
Note that if the breakdown occurs when the machine is processing a product, the product will be damaged and cannot be recovered.
The rescheduling decision-making considers the completion time of $s_{new}$ as the objective $C$, and this case study does not have soft constraints, thus $\alpha=1$ and $p_i=0$.
We conduct two case studies to evaluate the performance of the proposed distributed method.

\subsubsection{Centralized versus distributed}
The first case study aims to evaluate the performance trade-offs between the centralized method and the proposed distributed method in terms of optimal cycle time and computational efforts.
We run two simulation scenarios where the system uses centralized and distributed methods respectively as the rescheduling decision-making strategy. 
Note that risks are not included in this case study since it does not affect these trade-offs.
For each scenario, we run 5 trials to evaluate their performance with the following metrics:
\begin{itemize}
    \item Cycle time
    \item Number of agent communications
    \item Running time of the decision-making implementation
\end{itemize}
The number of communications in the centralized method includes the request for rescheduling, the requests to and responses from all the RAs in the system to collect information, and the notifications to the agents whose production schedules need to change.
Therefore, each rescheduling process requires $r\times\sum_{e_{d_i}\in E_d}|s_{d_i}|$ communications, where $r$ is the number of RAs in the system.
In the distributed method, the number of communication includes all agent requests, responses, and inform messages, as defined in Section~IV. 
The communication only occurs within local clustering RAs and their collaborative RAs (i.e., a subset of all the RAs in the system), thus the distributed method requires less communication, as showcased in Table~\ref{tab:result}.

\begin{figure}[tb]
    \centering
    \includegraphics[width=1\columnwidth]{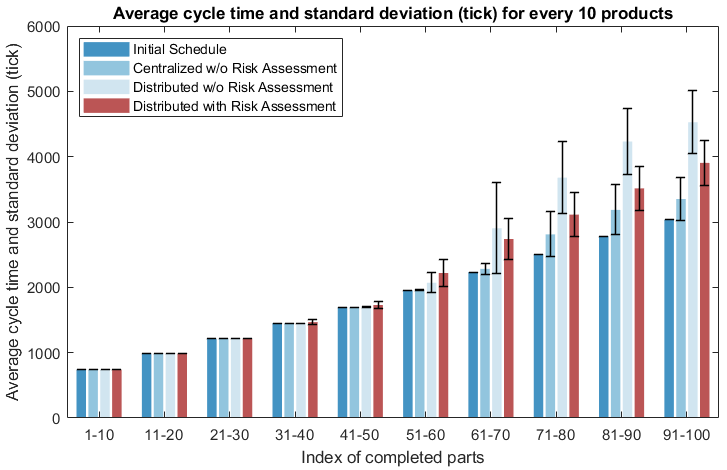}
    \caption{Average cycle time and standard deviation for every 10 products for 5 trials in different scenarios}
    \label{fig:meanth}
\end{figure}

Figure~\ref{fig:meanth} shows the product cycle time in different scenarios.
The centralized method re-optimizes the whole system to generate a new schedule with a shorter cycle time than the distributed method.
However, as shown in the first part of Table~\ref{tab:result}, the centralized method requires more communication and larger computational efforts to reschedule the system.
In practice, more communication potentially leads to a larger information delay, thus the centralized method lacks the ability to respond to disruptions dynamically and quickly.
Furthermore, the computational efforts of the centralized method increase as the size and complexity of the set-up increase in scale.
In this case, the distributed method can provide advantages by using local communication to reduce the communication and computation time.


\subsubsection{With risk assessment versus without risk assessment}

\begin{table*}[t]
    \caption{Evaluation of Performance Metrics for Different Trials}
        \label{tab:result}
        \centering
        \begin{tabularx}{1.95\columnwidth}{c@{\hskip 3em}c@{\hskip 3em}c@{\hskip 2em}c@{\hskip 2em}c@{\hskip 2em}c@{\hskip 2em}c@{\hskip 3em}c@{\hskip 2em}c}
        \hline
           \textbf{Metrics}  &\textbf{Scenarios}  &  \multicolumn{5}{c}{\textbf{Values in 5 trials}} & \textbf{Average}& \textbf{Percentage}\\
           \hline
           \noalign{\vskip 2pt}
          \multicolumn{9}{c}{\textbf{Centralized versus distributed}}\\
          \hline
          \noalign{\vskip 2pt}
          \multirow{2}{12em}{\# of agent communication} & Centralized  & 1430  & 1155 & 1650 & 1100 & 1320 & 1331 & N/A\\
           & Distributed  & 1053 & 858 & 1248 & 780 & 1014 & 991 & N/A\\ 
           \hline
           \noalign{\vskip 2pt}
           \multirow{2}{12em}{Total running time (sec) of rescheduling processes}& Centralized  & 10.1 & 18.7 & 16.4 & 8.8 & 9.4 & 12.7 & N/A\\
           & Distributed  & 0.22 & 0.18 & 0.34 & 0.49 & 0.33 & 0.31 & N/A\\ 
           \hline
           \noalign{\vskip 2pt}
          \multicolumn{9}{c}{\textbf{With risk assessment versus without risk assessment}}\\
          \hline
          \noalign{\vskip 2pt}
           \multirow{2}{10em}{\# of damaged products}&W/ risk assessment  & 3 & 5 & 3 & 3 & 5 & 3.8 &3.8\%\\
           &W/o risk assessment  & 9 & 7 & 8 & 6 & 5 & 7.0 & 7.0\%\\ 
           \hline
           \noalign{\vskip 2pt}
           \multirow{2}{10em}{\# of broken machines}&W/ risk assessment  & 6 & 7 & 4 & 5 & 7 & 5.8 &29\%\\
           &W/o risk assessment  & 11 & 10 & 10 & 8 & 9 & 9.6 & 48\%\\ 
           \hline
           \noalign{\vskip 2pt}
           \multirow{2}{12em}{\# of rescheduled processes}&W/ risk assessment  & 17 & 20 & 7 & 14 & 23 & 16.2 & 4.05\%\\
           &W/o risk assessment  & 32 & 25 & 35 & 21 & 31 & 28.8 & 7.20\%\\ 
           \hline
           \noalign{\vskip 2pt}
           \multirow{2}{10em}{Peak risk values}&W/ risk assessment  & 0.26&	0.26&	0.27&	0.26&	0.27 & 0.27 &N/A \\
           &W/o risk assessment  & 0.27&	0.29&	0.28&	0.29&	0.29& 0.28 &N/A\\ 
           \hline
           \noalign{\vskip 2pt}
           \multirow{2}{10em}{Average risk values}&W/ risk assessment  & 0.23&	0.22&	0.21&	0.22&	0.22 & 0.22 &N/A \\
           &W/o risk assessment  & 0.25&	0.26&	0.24&	0.24&	0.25& 0.25 &N/A\\ 
           \hline
        \end{tabularx}
\end{table*}
To further investigate how the introduction of risk assessment affects the rescheduling decision-making and the system performance, we run 5 trials where the system uses the distributed method with the incorporation of risk assessment into the rescheduling decision-making process.
As discussed in Section~V, the overall risk associated with a new schedule is calculated as the weighted sum of two risk factors, $w_1*R_1 + w_2*R_2$. 
In this case study, we simulated a scenario where the manufacturer cares more about machine breakdowns, thus we selected $w_1=0.2$ and $w_2=0.8$.
Note that different risks and weights can be defined and chosen, while the performance under different parameters can be investigated in future work.
The weighting gain, $W$, for the total risk value has been selected to assign value to the introduction of risk and scaled to ensure comparable unit values.
Note that for this example, the mobile robots are assumed to be reliable and are not considered within the risk assessment at this time.




Besides cycle time, we introduce the following metrics to evaluate the system performance with risk assessment:
\begin{itemize}
    \item Number of damaged products
    \item Number of broken machines
    \item Number of rescheduled processes
    \item Peak and average risk values of the new schedule
\end{itemize}

The results are shown in the second part of Table~\ref{tab:result}, which indicates that system performance with and without risk assessment varies across the different trials.
With risk assessment, the average number of damaged products is 3.8 versus an average number of 7 without risk assessment. When combined with the results from the number of machine breakdowns (5.8 versus 9.6), these results illustrate how the consideration of risk results in a rescheduling strategy that selects a less risky schedule that reduces the potential for machine breakdowns and damaged products.
Note that breakdowns may occur while the machines are not processing, thus the number of broken machines is larger than the number of damaged products.

To investigate how the rescheduling strategy impacts the potential for machine breakdown and the trigger of a new rescheduling task, the number of rescheduled processes is also presented in Table.~\ref{tab:result}.
On average, when risk assessment is included, the rescheduling process is triggered 16.2 times, while it is triggered 28.8 times when risk assessment is ignored.
Peak and average risk values provide a measure of the associated risks inherent in the two strategies. Note that when risk is included in the cost function, the decision-making strategy results in a selection process that chooses the event sequences with lower risks (0.27 versus 0.28 peak and 0.22 versus 0.25 average risk values).




To show how the assessment of risk affects the completion of products within the simulated facility, the mean values and the standard deviations of the average cycle time for every 10 products for the 5 trials are shown in Fig.~\ref{fig:meanth}.
As Fig.~\ref{fig:meanth} shows, the first 40 products have nearly identical cycle times. Although these products might be in the system during a later breakdown event, the risk of machine breakdown and a rescheduling event is low during this initial period. 

Interestingly, the impact of risk assessment really becomes apparent during the 61-70 part completion set. At this point in the simulation, the risk for machine breakdown is increasing as machine usage time gets closer to the MTBF for a given resource. Once R2 begins to increase, the decision to select the less risky event sequence results in fewer machine breakdowns, less rescheduling, and a lower average cycle time. This trend continues, with considerable variability beginning to be introduced into the cycle times as illustrated in Fig.~\ref{fig:meanth}.


Overall, the consideration of risk into the event sequence decision results in fewer damaged products and broken machines, a reduction in the number of triggered rescheduling processes, and an improvement in the system throughput as compared to the decision strategy that does not consider risk. 
These results showcased that incorporating risk assessment affects the agent decision-making in the rescheduling process.
Our results indicate that the introduction of the risk assessment value resulted in a smaller number of broken machines and damaged products, as well as a reduction in cycle time variability.   

\subsection{Insights from the case study}

The case study has showcased the feasibility and performance of the proposed multi-agent framework, specifically demonstrating how risk assessment affects agent decision-making.
However, there are other aspects that may affect the framework performance, which will be investigated in future work.
Firstly, the framework can be easily adapted to different case study setups.
As the set-up scales down, the set of candidate solutions might shrink.
Thus, there might be no big difference between centralized and distributed methods in terms of needed communications.
Besides, the risk assessment may not affect decision-making significantly since the choices are limited.
On the other hand, as the size and complexity of the set-up increase in scale, both the size and the variety of the candidate solutions might grow.
Therefore, the risk assessment can make a big difference in the selection of the new schedule.

Based on the objectives, risks, and parameters used in the simulation, this case study simulated a scenario where the manufacturer cares about cycle time and machine breakdown.
The results indicate that the proposed method reduced the production cycle time and machine breakdowns, which showcased the feasibility and performance of the proposed method.
Therefore, different objectives, risks, parameters, and metrics can be used in the proposed method, while it inevitably might change the results toward a better or worse direction.
An enhanced understanding of the sensitivity of the parameters design, such as identifying the set of conditions under which the algorithm always outperforms other algorithms, is left for future work.

In addition, other metrics, such as makespan and machine utilization rate can also be analyzed based on the provided results.
The cycle time in Fig.~\ref{fig:meanth} can reflect the makespan since the entry time of a specific product in each trial is identical.
Therefore, without risk assessment, the makespan of the schedule is larger.
Note that the large makespan occurs in the case where the system produced fewer products.
As a result, the machine utilization rate when the rescheduling does not consider risks is lower than the scenario when the risk assessment is incorporated.

From the managerial perspective, manufacturers can use this work to model and monitor their factory floor as the agents store the physical information and keep it updated.
Furthermore, this framework can be used as a decision support system since the agent decision-making ability can provide manufacturers with several solutions to respond to disruption depending on different objectives and parameters defined by the manufacturers.

\section{Conclusion}
\label{sec:conclusion}
Various multi-agent frameworks have been proposed to solve the dynamic rescheduling problem in manufacturing systems, where a resource agent (RA) is an important component in the existing multi-agent frameworks.
In this paper, a model-based RA architecture that enables effective agent coordination and dynamic decision-making is designed.
The proposed RA architecture contains a Knowledge Base, Decision Manager, and Communication Manager.
Based on this architecture, this paper developed a rescheduling strategy that incorporates risk assessment via RA coordination in the presence of resource breakdown.
The proposed work can be used to create manufacturing resource models that enable dynamic and resilient rescheduling for manufacturing systems. 
Implementation of the proposed framework in a simulation-based case study has been done to evaluate the effectiveness of the proposed architecture. 
In particular, the case study demonstrates that 
the proposed agent-based distributed method reduces the communications and computational efforts that are needed for rescheduling while losing some optimality in throughput compared to the centralized method.
Additionally, the case study illustrates the improvement in throughput when risk is considered within the rescheduling problem.
Future work may explore the advantages of a combined centralized and distributed framework to reduce the rescheduling effort.
In addition, learning algorithms may be used to incorporate historical information into the agent knowledge and risk assessment.

\ifCLASSOPTIONcaptionsoff
  \newpage
\fi


\begin{thebibliography}{10}
\providecommand{\url}[1]{#1}
\csname url@samestyle\endcsname
\providecommand{\newblock}{\relax}
\providecommand{\bibinfo}[2]{#2}
\providecommand{\BIBentrySTDinterwordspacing}{\spaceskip=0pt\relax}
\providecommand{\BIBentryALTinterwordstretchfactor}{4}
\providecommand{\BIBentryALTinterwordspacing}{\spaceskip=\fontdimen2\font plus
\BIBentryALTinterwordstretchfactor\fontdimen3\font minus
  \fontdimen4\font\relax}
\providecommand{\BIBforeignlanguage}[2]{{%
\expandafter\ifx\csname l@#1\endcsname\relax
\typeout{** WARNING: IEEEtran.bst: No hyphenation pattern has been}%
\typeout{** loaded for the language `#1'. Using the pattern for}%
\typeout{** the default language instead.}%
\else
\language=\csname l@#1\endcsname
\fi
#2}}
\providecommand{\BIBdecl}{\relax}
\BIBdecl

\bibitem{kumar2020covid}
A.~Kumar, S.~Luthra, S.~K. Mangla, and Y.~Kazan{\c{c}}o{\u{g}}lu, ``Covid-19
  impact on sustainable production and operations management,''
  \emph{Sustainable Operations and Computers}, vol.~1, pp. 1--7, 2020.

\bibitem{bi2022model}
M.~Bi, G.~Chen, D.~M. Tilbury, S.~Shen, and K.~Barton, ``A model-based
  multi-agent framework to enable an agile response to supply chain
  disruptions,'' in \emph{2022 IEEE 18th International Conference on Automation
  Science and Engineering (CASE)}.\hskip 1em plus 0.5em minus 0.4em\relax IEEE,
  2022, pp. 235--241.

\bibitem{li2020intelligent}
X.~Li, B.~Wang, C.~Liu, T.~Freiheit, and B.~I. Epureanu, ``Intelligent
  manufacturing systems in covid-19 pandemic and beyond: framework and impact
  assessment,'' \emph{Chinese Journal of Mechanical Engineering}, vol.~33,
  no.~1, pp. 1--5, 2020.

\bibitem{abumaizar1997rescheduling}
R.~J. Abumaizar and J.~A. Svestka, ``Rescheduling job shops under random
  disruptions,'' \emph{International journal of production research}, vol.~35,
  no.~7, pp. 2065--2082, 1997.

\bibitem{lee2022reinforcement}
J.-H. Lee and H.-J. Kim, ``Reinforcement learning for robotic flow shop
  scheduling with processing time variations,'' \emph{International Journal of
  Production Research}, vol.~60, no.~7, pp. 2346--2368, 2022.

\bibitem{liu2022deep}
R.~Liu, R.~Piplani, and C.~Toro, ``Deep reinforcement learning for dynamic
  scheduling of a flexible job shop,'' \emph{International Journal of
  Production Research}, pp. 1--21, 2022.

\bibitem{park2019reinforcement}
I.-B. Park, J.~Huh, J.~Kim, and J.~Park, ``A reinforcement learning approach to
  robust scheduling of semiconductor manufacturing facilities,'' \emph{IEEE
  Transactions on Automation Science and Engineering}, vol.~17, no.~3, pp.
  1420--1431, 2019.

\bibitem{yang2021intelligent}
S.~Yang and Z.~Xu, ``Intelligent scheduling and reconfiguration via deep
  reinforcement learning in smart manufacturing,'' \emph{International Journal
  of Production Research}, pp. 1--18, 2021.

\bibitem{palombarini2021end}
J.~A. Palombarini and E.~C. Mart{\'\i}nez, ``End-to-end on-line rescheduling
  from gantt chart images using deep reinforcement learning,''
  \emph{International Journal of Production Research}, pp. 1--30, 2021.

\bibitem{leitao2009agent}
P.~Leit{\~a}o, ``Agent-based distributed manufacturing control: A
  state-of-the-art survey,'' \emph{Engineering applications of artificial
  intelligence}, vol.~22, no.~7, pp. 979--991, 2009.

\bibitem{shen2006agent}
W.~Shen, L.~Wang, and Q.~Hao, ``Agent-based distributed manufacturing process
  planning and scheduling: a state-of-the-art survey,'' \emph{IEEE Transactions
  on Systems, Man, and Cybernetics, Part C (Applications and Reviews)},
  vol.~36, no.~4, pp. 563--577, 2006.

\bibitem{kovalenko2022toward}
I.~Kovalenko, J.~Moyne, M.~Bi, E.~C. Balta, W.~Ma, Y.~Qamsane, X.~Zhu, Z.~M.
  Mao, D.~M. Tilbury, and K.~Barton, ``Toward an automated learning control
  architecture for cyber-physical manufacturing systems,'' \emph{IEEE Access},
  vol.~10, pp. 38\,755--38\,773, 2022.

\bibitem{bi2021dynamic}
M.~Bi, I.~Kovalenko, D.~M. Tilbury, and K.~Barton, ``Dynamic resource
  allocation using multi-agent control for manufacturing systems,''
  \emph{IFAC-PapersOnLine}, vol.~54, no.~20, pp. 488--494, 2021.

\bibitem{wooldridge2009introduction}
M.~Wooldridge, \emph{An introduction to multiagent systems}.\hskip 1em plus
  0.5em minus 0.4em\relax John wiley \& sons, 2009.

\bibitem{kovalenko2019dynamic}
I.~Kovalenko, D.~Ryashentseva, B.~Vogel-Heuser, D.~Tilbury, and K.~Barton,
  ``Dynamic resource task negotiation to enable product agent exploration in
  multi-agent manufacturing systems,'' \emph{IEEE Robotics and Automation
  Letters}, vol.~4, no.~3, pp. 2854--2861, 2019.

\bibitem{kovalenko2019model}
I.~Kovalenko, D.~Tilbury, and K.~Barton, ``The model-based product agent: A
  control oriented architecture for intelligent products in multi-agent
  manufacturing systems,'' \emph{Control Engineering Practice}, vol.~86, pp.
  105--117, 2019.

\bibitem{mejia2020robust}
G.~Mej{\'\i}a and D.~Lefebvre, ``Robust scheduling of flexible manufacturing
  systems with unreliable operations and resources,'' \emph{International
  Journal of Production Research}, vol.~58, no.~21, pp. 6474--6492, 2020.

\bibitem{zhang2017flexible}
S.~Zhang and T.~N. Wong, ``Flexible job-shop scheduling/rescheduling in dynamic
  environment: a hybrid mas/aco approach,'' \emph{International Journal of
  Production Research}, vol.~55, no.~11, pp. 3173--3196, 2017.

\bibitem{rehberger2016agent}
S.~Rehberger, L.~Spreiter, and B.~Vogel-Heuser, ``An agent approach to flexible
  automated production systems based on discrete and continuous reasoning,'' in
  \emph{2016 IEEE International Conference on Automation Science and
  Engineering (CASE)}.\hskip 1em plus 0.5em minus 0.4em\relax IEEE, 2016, pp.
  1249--1256.

\bibitem{wong2006integrated}
T.~Wong, C.~Leung, K.~Mak, and R.~Fung, ``Integrated process planning and
  scheduling/rescheduling—an agent-based approach,'' \emph{International
  Journal of Production Research}, vol.~44, no. 18-19, pp. 3627--3655, 2006.

\bibitem{rodrigues2018decentralized}
N.~Rodrigues, E.~Oliveira, and P.~Leit{\~a}o, ``Decentralized and on-the-fly
  agent-based service reconfiguration in manufacturing systems,''
  \emph{Computers in Industry}, vol. 101, pp. 81--90, 2018.

\bibitem{fu2020heterogeneous}
B.~Fu, W.~Smith, D.~Rizzo, M.~Castanier, and K.~Barton, ``Heterogeneous vehicle
  routing and teaming with gaussian distributed energy uncertainty,'' in
  \emph{2020 IEEE/RSJ International Conference on Intelligent Robots and
  Systems (IROS)}.\hskip 1em plus 0.5em minus 0.4em\relax IEEE, 2020, pp.
  4315--4322.

\bibitem{fu2022robust}
B.~Fu, W.~Smith, D.~M. Rizzo, M.~Castanier, M.~Ghaffari, and K.~Barton,
  ``Robust task scheduling for heterogeneous robot teams under capability
  uncertainty,'' \emph{IEEE Transactions on Robotics}, 2022.

\bibitem{lepuschitz2010toward}
W.~Lepuschitz, A.~Zoitl, M.~Vall{\'e}e, and M.~Merdan, ``Toward
  self-reconfiguration of manufacturing systems using automation agents,''
  \emph{IEEE Transactions on Systems, Man, and Cybernetics, Part C
  (Applications and Reviews)}, vol.~41, no.~1, pp. 52--69, 2010.

\bibitem{uhlmann2022hybrid}
I.~R. Uhlmann, R.~M. Zanella, and E.~M. Frazzon, ``Hybrid flow shop
  rescheduling for contract manufacturing services,'' \emph{International
  Journal of Production Research}, vol.~60, no.~3, pp. 1069--1085, 2022.

\bibitem{antzoulatos2017multi}
N.~Antzoulatos, E.~Castro, L.~de~Silva, A.~D. Rocha, S.~Ratchev, and J.~Barata,
  ``A multi-agent framework for capability-based reconfiguration of industrial
  assembly systems,'' \emph{International Journal of Production Research},
  vol.~55, no.~10, pp. 2950--2960, 2017.

\bibitem{vieira2003rescheduling}
G.~E. Vieira, J.~W. Herrmann, and E.~Lin, ``Rescheduling manufacturing systems:
  a framework of strategies, policies, and methods,'' \emph{Journal of
  scheduling}, vol.~6, no.~1, pp. 39--62, 2003.

\bibitem{farid2015axiomatic}
A.~M. Farid and L.~Ribeiro, ``An axiomatic design of a multiagent
  reconfigurable mechatronic system architecture,'' \emph{IEEE Transactions on
  Industrial Informatics}, vol.~11, no.~5, pp. 1142--1155, 2015.

\bibitem{park2012autonomous}
H.-S. Park and N.-H. Tran, ``An autonomous manufacturing system based on swarm
  of cognitive agents,'' \emph{Journal of Manufacturing Systems}, vol.~31,
  no.~3, pp. 337--348, 2012.

\bibitem{maturana1999metamorph}
F.~Maturana, W.~Shen, and D.~H. Norrie, ``Metamorph: an adaptive agent-based
  architecture for intelligent manufacturing,'' \emph{International Journal of
  Production Research}, vol.~37, no.~10, pp. 2159--2173, 1999.

\bibitem{barata2003coalitions}
J.~Barata and L.~M. Camarinha-Matos, ``Coalitions of manufacturing components
  for shop floor agility-the cobasa architecture,'' \emph{International journal
  of networking and virtual organisations}, vol.~2, no.~1, pp. 50--77, 2003.

\bibitem{kim2020multi}
Y.~G. Kim, S.~Lee, J.~Son, H.~Bae, and B.~Do~Chung, ``Multi-agent system and
  reinforcement learning approach for distributed intelligence in a flexible
  smart manufacturing system,'' \emph{Journal of Manufacturing Systems},
  vol.~57, pp. 440--450, 2020.

\bibitem{lee2013risk}
C.~Lee, Y.~Lv, and Z.~Hong, ``Risk modelling and assessment for distributed
  manufacturing system,'' \emph{International Journal of Production Research},
  vol.~51, no.~9, pp. 2652--2666, 2013.

\bibitem{anbarani2022risk}
M.~T. Anbarani, E.~C. Balta, R.~Meira-G{\'o}es, and I.~Kovalenko, ``Risk-averse
  model predictive control for priced timed automata,'' \emph{arXiv preprint
  arXiv:2210.15604}, 2022.

\bibitem{liu2020parallel}
X.~Liu, F.~Chu, F.~Zheng, C.~Chu, and M.~Liu, ``Parallel machine scheduling
  with stochastic release times and processing times,'' \emph{International
  Journal of Production Research}, pp. 1--20, 2020.

\bibitem{klober2017predictive}
J.~Kl{\"o}ber-Koch, S.~Braunreuther, and G.~Reinhart, ``Predictive production
  planning considering the operative risk in a manufacturing system,''
  \emph{Procedia CIRP}, vol.~63, pp. 360--365, 2017.

\bibitem{guo2021sequencing}
G.~Guo and S.~M. Ryan, ``Sequencing mixed-model assembly lines with risk-averse
  stochastic mixed-integer programming,'' \emph{International Journal of
  Production Research}, pp. 1--18, 2021.

\bibitem{framinan2019using}
J.~M. Framinan, V.~Fernandez-Viagas, and P.~Perez-Gonzalez, ``Using real-time
  information to reschedule jobs in a flowshop with variable processing
  times,'' \emph{Computers \& Industrial Engineering}, vol. 129, pp. 113--125,
  2019.

\bibitem{toothman2021trend}
M.~Toothman, B.~Braun, S.~J. Bury, M.~Dessauer, K.~Henderson, R.~Wright, D.~M.
  Tilbury, J.~Moyne, and K.~Barton, ``Trend-based repair quality assessment for
  industrial rotating equipment,'' in \emph{2021 American Control Conference
  (ACC)}.\hskip 1em plus 0.5em minus 0.4em\relax IEEE, 2021, pp. 502--507.

\bibitem{howden2001jack}
N.~Howden, R.~R{\"o}nnquist, A.~Hodgson, and A.~Lucas, ``Jack intelligent
  agents-summary of an agent infrastructure,'' in \emph{5th International
  conference on autonomous agents}, vol. 142, 2001.

\bibitem{kovalenko2022cooperative}
I.~Kovalenko, E.~C. Balta, D.~M. Tilbury, and K.~Barton, ``Cooperative product
  agents to improve manufacturing system flexibility: A model-based decision
  framework,'' \emph{IEEE Transactions on Automation Science and Engineering},
  2022.

\bibitem{qamsane2019dynamic}
Y.~Qamsane, E.~C. Balta, J.~Moyne, D.~Tilbury, and K.~Barton, ``Dynamic
  rerouting of cyber-physical production systems in response to disruptions
  based on sdc framework,'' in \emph{2019 American Control Conference
  (ACC)}.\hskip 1em plus 0.5em minus 0.4em\relax IEEE, 2019, pp. 3650--3657.

\bibitem{balta2021model}
E.~C. Balta, I.~Kovalenko, I.~A. Spiegel, D.~M. Tilbury, and K.~Barton, ``Model
  predictive control of priced timed automata encoded with first-order logic,''
  \emph{IEEE Transactions on Control Systems Technology}, 2021.

\bibitem{macal2006introduction}
C.~M. Macal and M.~J. North, ``Introduction to agent-based modeling and
  simulation,'' in \emph{Proceedings of the MCS LANS Informal Seminar}, 2006.

\bibitem{yoon2008multiagent}
H.~J. Yoon and W.~Shen, ``A multiagent-based decision-making system for
  semiconductor wafer fabrication with hard temporal constraints,'' \emph{IEEE
  Transactions on Semiconductor Manufacturing}, vol.~21, no.~1, pp. 83--91,
  2008.

\end{thebibliography}

\begin{thebibliography}{1}

\bibitem{IEEEhowto:kopka}
H.~Kopka and P.~W. Daly, \emph{A Guide to \LaTeX}, 3rd~ed.\hskip 1em plus
  0.5em minus 0.4em\relax Harlow, England: Addison-Wesley, 1999.

\end{thebibliography}
\end{document}